\documentclass[10pt,journal,compsoc]{IEEEtran}
\pdfoutput=1
\usepackage{xspace}
\usepackage[dvipsnames]{color}

\newcommand{\subparagraph}{}
\usepackage{graphicx}
\usepackage{amsmath}
\usepackage{array}
\usepackage{enumitem}
\usepackage{epsfig}
\usepackage{color}
\usepackage{subfigure} 
\usepackage{spverbatim}
\usepackage{gensymb}
\usepackage{algorithm,algorithmic}
\usepackage{amsmath}
\usepackage{amsfonts}
\usepackage{amssymb}
\usepackage{multirow}
\usepackage{etoolbox}
\usepackage{caption}

\DeclareMathOperator*{\argmax}{argmax}
\newcommand{\wifi}{WiFi\xspace}
\newcommand{\mquote}[1]{``#1''} 

\makeatletter
\patchcmd{\@maketitle}
  {\addvspace{0.5\baselineskip}\egroup}
  {\addvspace{-1.0\baselineskip}\egroup}
  {}
  {}
\makeatother

\begin{document}
\title{Handcrafted vs Deep Learning Classification for Scalable Video QoE Modeling}
\author{\IEEEauthorblockN{Mallesham Dasari\IEEEauthorrefmark{4}, Christina Vlachou\IEEEauthorrefmark{2}, Shruti Sanadhya\IEEEauthorrefmark{2}\IEEEauthorrefmark{3}, Pranjal Sahu\IEEEauthorrefmark{4}, Yang Qiu\IEEEauthorrefmark{4}, \\ Kyu-Han Kim\IEEEauthorrefmark{2}, Samir R. Das\IEEEauthorrefmark{4}} \\
    \IEEEauthorblockA{\IEEEauthorrefmark{4}Stony Brook University, \IEEEauthorrefmark{2}HPE Labs, \IEEEauthorrefmark{3}Facebook}
    }

\maketitle
\begin{abstract} 
Mobile video traffic is dominant in cellular and enterprise wireless networks.
With the advent of diverse applications, network administrators face the challenge to provide high QoE in the face of diverse wireless conditions and application contents. Yet, state-of-the-art networks lack analytics for QoE, as this requires support from the application or user feedback.
While there are existing techniques to map QoS to QoE by training machine learning models without requiring user feedback, these techniques are limited to only few applications, due to insufficient QoE ground-truth annotation for ML. 
To address these limitations, we focus on video telephony applications and model key artefacts of spatial and temporal video QoE. 
Our key contribution is designing content- and device-independent metrics and training across diverse WiFi conditions. We show that our metrics achieve a median 90\% accuracy 
by comparing with mean-opinion-score from more than 200 users and 800 video samples over three popular video telephony applications -- Skype, FaceTime and Google Hangouts.
We further extend our metrics by using deep neural networks, more specifically we use a combined CNN and LSTM model.
We achieve a median accuracy of 95\% by combining our QoE metrics with the deep learning model, which is a 38\% improvement over the state-of-the-art well known techniques. 
\end{abstract}

\section{Introduction}


Over the past decade, mobile video traffic has increased dramatically (from 50\% in year 2011 to 60\% in 2016 and is predicted to reach 78\% by 2021)~\cite{forecast2016cisco}. This is due to the proliferation of mobile video applications (such as Skype, FaceTime, Hangouts, YouTube, and Netflix etc). These applications can be categorized into video telephony (Skype, Hangouts, FaceTime), streaming (YouTube, Netflix), and upcoming virtual reality and augmented reality streaming (SPT~\cite{spt}, theBlu~\cite{theblu}). 
Users demand high Quality of Experience (QoE) while using these applications on wireless networks, such as \wifi and LTE. This poses a unique challenge for network administrators in enterprise environments, such as offices, university campuses and retail stores.

Guaranteeing best possible QoE is non-trivial because of several factors in video delivery path (such as network conditions at the client side and server side, client device, video compression standard, video application). While application content providers focus on improving the server-side network performance, video compression and application logic, enterprise network administrators seek to ensure that network resources are well provisioned and managed to provide good experience to multiple users of diverse applications. In this pursuit, network administrators can only rely on passive in-network measurements to {\em estimate} the exact user experience on the end-device. In this context, several prior works have developed a mapping between network Quality-of-Service (QoS) and end-user QoE for video applications~\cite{balachandran2013developing}\cite{aggarwal2014prometheus}\cite{fiedler2010generic}, by using machine learning (ML) models and network features from PHY to TCP/UDP layers. ML models for QoE can be deployed at a number of vantage points in the network, such as access points, network controllers, cellular packet core. 

In order to train any QoS to QoE model, one needs accurate ground-truth annotation of the QoE. To obtain this, prior work has leveraged application specific APIs such as Skype technical information~\cite{zhang2012profiling} or YouTube API~\cite{aggarwal2014prometheus}; call duration~\cite{chen2006quantifying}; or instrumented client-side libraries for video delivery platform~\cite{balachandran2013developing}. While these solutions perform well for specific applications and providers, network administrators have to \emph{deal with a plethora of video applications} and they \emph{cannot control the application logic} for most applications. Nor does every application expose QoE metrics through APIs. 
Thus, to develop QoS to QoE models in the wild, network administrators need an application-independent approach to measure QoE. Note that application independence does not mean that modeling is application-agnostic, as a single QoE model cannot apply to all applications. Different applications exhibit diverse artefacts when quality deteriorates. For instance, streaming quality is impacted largely by buffering and stall ratio, whereas telephony quality is impacted by bit rate, frames per second, blocking and blurring in the video.

In this work, we propose a generic video telephony QoE model by leveraging existing machine/deep learning models, that does not rely on application support. Additionally, it is scalable to diverse content, devices and categories of video application. We exploit video compression methods (Section~\ref{label:background}) and identify four new metrics: {\em perceptual bitrate (PBR)}, {\em freeze ratio}, {\em length} and {\em number of video freezes} to measure the video QoE. We further demonstrate that our metrics are insensitive to the content of the video call and they only capture the {\em quality} of the video call (Section~\ref{label:results}). We make a rigorous analysis of our model performance by conducting a large scale user-study of 800 video clips across 20 videos and more than 200 users. We conduct an extensive analysis across different devices and OS (Android vs. iOS), video content and motion.  

We also show that these metrics capture spatial and temporal artefacts of video experience.
We then validate our metrics by mapping them to actual users' experience. To this end, we obtain Mean Opinion Score (MOS) from our user-study and apply ML models to map our metrics with users' MOS.
We use \texttt{Adaboosted} decision trees in predicting the MOS scores, as we describe in Section \ref{label:results}.

This paper is an extension of our previous work \cite{mdasariiwqos2018} that models the QoE of end users using the handcrafted features described above. 
We extend this conference version by micro-benchmarking our metrics \cite{mdasariiwqos2018} under different network conditions to show that the metrics are agnostic to motion and content of the video.
Second, we analyze the variation between use QoE ratings in different environments (e.g., Lab vs Amazon mechanical turk users). 
Third, to improve the accuracy, we seek recent success of deep learning solutions \cite{lecun2015deep}\cite{karpathy2014large}\cite{ng2015beyond} by formulating the video quality assessment as a classification problem. In particular, we use convolutional neural networks (CNNs) \cite{krizhevsky2012imagenet} to capture spatial artefacts and Long Short Term Memory (LSTM) network \cite{lstmcolah2018} to capture temporal artefacts of video. Finally, we propose a hybrid model that includes the features from deep neural networks and handcrafted features in classifying the video quality.

In summary, our contributions are the following:
\begin{itemize}
    \item We uncover limitations of existing work for QoE annotation of video telephony in enterprise networks (Section~\ref{MOTIVATION}).
    \item We introduce new QoE metrics for video telephony that are content, application and device independent (Section~\ref{label:design}).
    \item We micro-benchmark our metrics with the three most popular applications, i.e., Skype, FaceTime and Hangouts, and five different mobile devices (Section~\ref{label:design}).
    \item We develop a model to map our QoE metrics to MOS and demonstrate a median 90\% accuracy across applications and devices (Section~\ref{label:results}).
    \item We propose generic deep learning model combined with handcrafted features and show a median accuracy of 96\% which is 6\% improvement over handcrafted features.
\end{itemize}

\section{Motivation for Scalable QoE Annotation} \label{MOTIVATION}

In this section, we describe the need for new QoE metrics for enterprise networks and the limitations of existing work. 

\noindent {\bf Lack of QoE information from applications}:
While several works have motivated and addressed the problem of network-based QoE estimation~\cite{aggarwal2014prometheus}, little attention has been paid to the problem of collecting the QoE ground-truth. Most works have relied on application-specific information. This approach is effective for QoE optimizations by content providers, as they only focus on a single application~\cite{balachandran2013developing} or initial training of QoS to QoE model~\cite{aggarwal2014prometheus}. Nevertheless, administrators of access networks have to ensure good experience for a wide range of diverse applications being simultaneously used. 
Table~\ref{tab:table1} shows that not all popular video applications provide QoE information. 
Even for applications like Skype, availability of technical information depends on the version of the application and on the OS (e.g., no technical information for iOS). 

\begin{table}[htb!]
	\centering
		\scriptsize
			\vspace*{-0.5em}	
		\scalebox{1.1}{
			\begin{tabular}{l|c}
				\hline \hline
				\textbf{Application} & \textbf{Ground-truth} \\
				\hline
				\hline
				Skype & \checkmark *on some versions \\  \hline
				Hangouts/Duo & $\times$ \\ \hline
				FaceTime & $\times$ \\ \hline
				YouTube/Netflix & \checkmark \\ \hline 
			\end{tabular} 
		}
        \vspace*{0.2em}
		\hfill 
		\caption{Availability of QoE ground-truth}
		\label{tab:table1} 
\end{table}

To apply QoE estimation models in real networks, we need to remove the dependency of QoE metrics on application features, such as Skype technical information. One way to achieve this is to simply record the video as it plays on the mobile device and analyze this video for quality. 

\noindent {\bf Lack of scalable and reliable  QoE measures:}
Prior work on video quality evaluation leverages \emph{subjective and/or objective} metrics. Subjective metrics are measured with MOS collected through user surveys. They capture absolute QoE but are tedious to conduct and to scale. QoE metrics need to scale to thousands of videos in order to train models that map QoS to QoE. Alternatively, objective metrics can be computed from the video. Objective metrics are further classified in two categories: reference and no-reference based. A reference-based metric uses both sent and received video, and compares the quality of sent vs.~received frames. As it is challenging to retrieve and synchronize reference videos for telephony applications, no-reference based quality metrics are preferred. Jana {\em et al}~\cite{jana2016qoe} have proposed a no-reference metric for QoE estimation in Skype and Vtok. They record received videos for each of these mobile telephony applications and compute three no-reference metrics: {\em blocking, blurring} and {\em temporal variations}. Then, they combine these three metrics into one QoE metric by using MOS from subjective user study. Their study shows that blocking does not impact MOS of a video clip. While they show that their {\em blur} and {\em temporal variation} metrics correlate well with MOS, they do not evaluate these metrics over a wide range of clips. For example, one can wonder if {\em blur} is sensitive to the video content. 

\begin{figure}[t]
	\centering
	\vspace*{-2em}
	\includegraphics[width=0.45\textwidth]{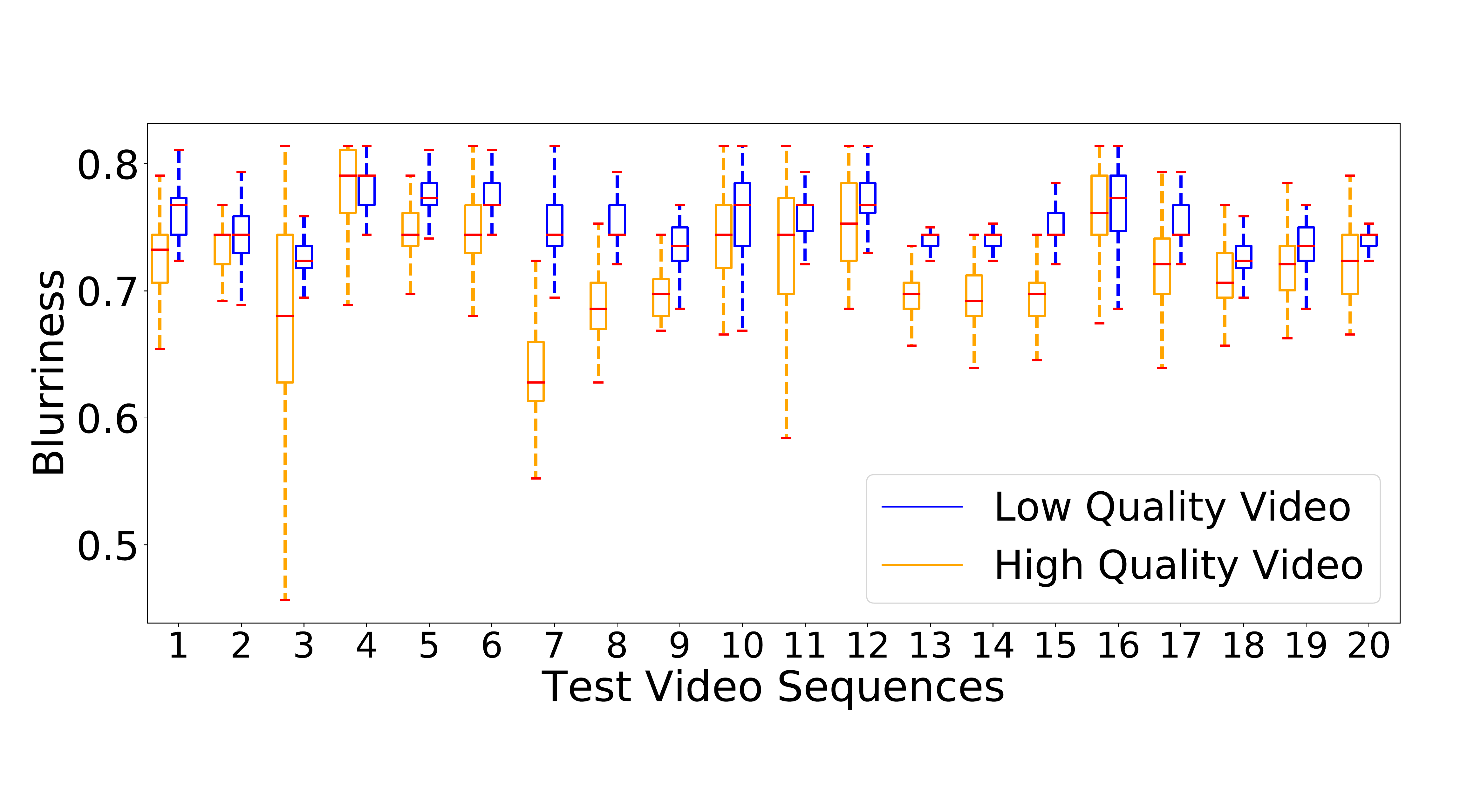}
	\vspace*{-2em}
	\caption{Blur detection using DCT coefficients~\cite{jana2016qoe,marichal1999blur}. DCT coefficients fail to distinguish between high and low quality video due to content diversity.}
	\label{fig:ucdavis}
	\vspace*{-0.5cm}
\end{figure}

We conduct similar experiments with Skype to evaluate blur metric of prior works. 
Our experimental setup is described in Section~\ref{label:design}. 
To capture video blur, Jana {\em et al}~\cite{jana2016qoe}  employ discrete cosine transform (DCT) coefficients \cite{marichal1999blur} from the compressed data by computing the histogram of DCT coefficients thereby characterizing the image quality. 
The DCT coefficients are obtained in transform coding of video compression process, as described in Section \ref{label:background}. 
The assumption here is that blurred image has high-frequency coefficients close to zero. 
Hence, the method studies the distribution of zero coefficients instead of absolute pixel values. 
Although the method estimates out-of-focus blur accurately, it falls short in estimating realistic blur and sensitivity to noise. 
The authors also point out that the method is very sensitive to uniform background and images with high luminance components. 
In Fig. \ref{fig:ucdavis}, we evaluate blur for 20 video sequences\footnote{\scriptsize{The videos are located at \textit{https://bit.ly/2CM1pEY}.}} using DCT metric. 
We have collected these videos in a representative manner to cover diverse content, different types of motion and we have downloaded them in Full HD resolution. 
The same 20 videos are converted to low quality by compressing and decreasing the resolution, to observe the difference of DCT metric between high and low quality videos.

We notice two aberrations of the DCT metric by Jana {\em et al}~\cite{jana2016qoe}: First, the metric is indeed content-specific i.e., although it produces high blur values for some low quality videos, it also shows high blur values for some high quality videos. For instance, videos 2 and 4 contain mostly over-illuminated and dark images and are equally tagged as blurred in both low and high quality scenarios. Second, DCT metric fails to detect accurate blur levels, even if the image is blurred heavily i.e., it shows low blur differences ($<$0.2 blurriness) between high and low quality videos even though we created extremely low-quality videos. Even for a single video, this DCT metric shows a lot of variation, such as for videos 3 and 11, raising concerns about its accuracy. We evaluate the accuracy of this blurriness metric in Section \ref{label:model}, showing a MOS error larger than 1.2. We also experiment with four other well-known blur metrics \cite{golestaneh2014no}\cite{mittal2012no}\cite{tong2004blur}\cite{marziliano2002no}, but none of these methods are consistent across diverse videos. This challenges the scalability and versatility of this blur metric in the wild. 

The {\em temporal variation} metric proposed by Jana {\em et al}~\cite{jana2016qoe} aims to capture video stalls and considers the ratio of missed frames to total number of frames in a video, but the metric requires the number of frames sent over the network. One needs the reference video to compute the number of frames in the video, thus the metric is not entirely no-reference. 
In enterprise networks, a QoE metric needs to be applicable to diverse contents and to not rely on access to reference video.
Further, other previous works \cite{wolf2009no}\cite{borer2010model}\cite{usman2017no}\cite{pastrana2006automatic} focused on measuring the temporal jerkiness. We find these methods  either are parametrized or are sensitive to resolution and to frame rate of the video or are unable to scale across diverse video contents. The above limitations motivate us to propose new metrics that can accurately measure blurriness and freezes across diverse video content.

\noindent {\bf Need for a model per application category:}
When analyzing a recorded video for an application, the QoE metric has to be sensitive to the application category as different applications have to meet diverse performance guarantees. For instance, streaming quality is impacted largely by buffering and stall ratio, whereas telephony quality is impacted by bit rate, frames per second, blocking and blurring in the video. Table~\ref{tab:qoe_metric} lists examples of QoE metrics corresponding to different applications. Therefore, one needs to capture different artefacts when designing models for multiple application categories.

To address the aforementioned requirements, we seek to answer the question: {\em How to scalably label the quality of a video call, without any support from the application?}

\begin{table}[htb!]
	\centering
	\scriptsize
	\vspace*{-0.5em}	
		\scalebox{1.1}{
			\begin{tabular}{l|c}				
				\hline \hline
				\textbf{Application Category} & \textbf{Suitable Metric} \\
				\hline
				\hline
				Adaptive Streaming & Startup delay, Buffering ratio, Stall ratio \\ \hline
				Video Telephony & Bitrate, Fps, Blocking, Blurring \\ \hline
				Progressive Downloads & PSNR, SSIM \\ \hline
				VR/AR, Game Streaming & Latency, buffering ratio, Stall ratio \\ \hline 
			\end{tabular}
		}
	\vspace*{0.5em}
		\hfill 
		\caption{Metrics based on application category}
		\label{tab:qoe_metric}
	\vspace*{-2em}
\end{table}

\section{Related Work}
There is an extensive prior work on video QoE modeling. The vast majority of works \cite{aggarwal2014prometheus}\cite{balachandran2013developing}\cite{fiedler2010generic} introduce machine learning methods to map QoS to QoE, and assume availability of the QoE ground-truth. Features for QoE estimation can be collected from various vantage points, such as on the application server~\cite{balachandran2013developing}\cite{balachandran2014modeling}\cite{dobrian2011understanding}, on the end-device (application-client statistics or packet capturing)~\cite{zhang2012profiling}\cite{yu2014can}\cite{seufert2015survey}\cite{aggarwal2014prometheus}\cite{dasari2018impact}, or the access network (e.g., \wifi AP or LTE base station)~\cite{chakraborty2016exbox}\cite{jana2016qoe}\cite{chen2006quantifying}. Compared to the above works, we focus on annotating QoE ground-truth, hence easing the extension of QoS to QoE mappings to all video-telephony applications. Our work significantly advances state-of-the-art QoE labeling at the training phase of QoS to QoE, by introducing accurate metrics that capture spatial and temporal video quality.

\noindent \textbf{Spatial quality assessment:} Prior works have proposed multiple metrics to capture video blurriness (spatial artefacts). Jana et al.~\cite{jana2016qoe} introduce metrics for freezes, blocking and blur over Skype and Vtok. However, as we show in Section~\ref{MOTIVATION}, this blur metric is highly content dependent. We have found similar results for major prior works~\cite{golestaneh2014no}\cite{mittal2012no}\cite{tong2004blur}\cite{marziliano2002no} on no-reference blur detection. In Section V, we present more than 1 MOS lower estimation error for our metrics compared to prior work over 20 diverse videos. We have not observed blocking artefacts in any of Skype, FaceTime or Hangouts applications.

\noindent \textbf{Temporal quality assessment:} Several works \cite{wolf2009no}\cite{borer2010model}\cite{usman2017no}\cite{pastrana2006automatic} have investigated the effect of video freezes on user QoE. Wolf and Pinson~\cite{wolf2009no} propose to use the motion energy temporal history of videos, along with framerate of video. This requires additional information from original video hence, making it a reduced reference metric. Similarly, the temporal metric of Jana et al.~\cite{jana2016qoe} is a reduced-reference metric. The temporal metric by Pastrana-Vidal and Gicquel~\cite{pastrana2006automatic} is sensitive to resolution and content of the video. Different from all of these works, we present three content-independent, no-reference temporal metrics to assess temporal video artefacts.

\noindent \textbf{Machine/Deep learning based quality assessment:} There has been a lot work in the space of image quality assessment using machine learning \cite{hou2015blind}\cite{ye2012unsupervised} and recently deep learning \cite{xue2013learning}\cite{narwaria2010objective}. A relatively few works have focused on video quality assessment \cite{narwaria2012low}. However, most of these models limited either spatial or temporal feature extraction. Moreover, the datasets used these models are not enough to evaluate the scalability of the models. In our model, we carefully select 20 different video as described in \S \ref{MOTIVATION} and apply both spatio-temporal feature extraction methods. There are also few deep learning models proposed to improve the QoE of video streaming \cite{mao2017neural}\cite{yeo2018neural}. However, these works are orthogonal to our work and focum more on rate adaptation algorithms.

\section{Background} \label{label:background}

Before designing our no-reference QoE metrics, we present a video coding primer and QoE artefacts in video telephony.
In Section~\ref{label:design}, we harness these coding properties to carefully craft accurate metrics that capture such artefacts.

\subsection{Video Coding Primer}
Video coding is performed in three critical steps: 

\begin{itemize}[leftmargin=*]
\item \textbf{Frame prediction:}
    The encoder takes images of video and divides each image into macroblocks (typically 16x16, 16x8, 8x16, 8x8, 4x4 pixel blocks). The macroblocks are predicted from previously encoded macroblocks, either from current image (called \emph{intra-frame prediction}) or from previous frames and future frames (called \emph{inter-frame prediction}). Depending on the intra- and inter-macroblocks, frames are classified as I, P and B frames. The I frame uses only intra-frame prediction i.e., it does not employ any previously coded frames as reference. Whereas P frames are predicted using previously coded frames and B frames are encoded from previous and future frames. The encoder typically combines both intra- and inter-prediction techniques to exploit spatial and temporal redundancy, respectively. Intra-frame prediction involves different prediction modes \cite{richardson2004h} while finding candidate macroblocks. Inter-frame prediction uses motion estimation by employing different block matching algorithms (such as Hexagon based or full exhaustive search) to identify the candidate macroblock across frames. Finally, a residual is calculated by taking the difference (measured typically using mean absolute difference (MAD) or mean squared error (MSE)) between the predicted and current macroblock. The prediction stage produces 4x4 to 16x16 blocks of absolute-pixel or residual values. 

\item \textbf{Transform coding and quantization:}
These absolute values are then transformed and quantized for further compression. Typically, two transform coding techniques (block or wavelet based) are used to convert pixel values into transform coefficients. A subset of these transform coefficients are sufficient to construct actual pixel values, which means reduced data upon quantization. The most popular transform coding is DCT over 8x8 macroblocks. 

\item \textbf{Entropy coding:}
Finally, coefficients are converted to binary data which is further compressed using entropy coding techniques, such as CABAC, CAVLC or Huffman. Richardson presents details of video compression~\cite{richardson2004h}.
\end{itemize}
We highlight that inter-frame prediction depends on motion content in the video. The higher the motion in the video is, the lower the compression rate is. Similarly, intra-frame compression is affected by frame contents such as color variation. For instance, the blurrier the video is, the higher the compression rate is. In the next section, we design QoE metrics exploiting these video coding principles.

\subsection{QoE Artefacts in Video Telephony}

When a user is in a video telephony call, she can experience different aberrations in video quality, as follows.

\noindent \textbf{Video freeze} is a temporal disruption in a video. A user may experience freeze when the incoming video stalls abruptly and the same frame is displayed for a long duration. Additionally, freeze may appear as a fast play of video, where the decoder tries to recover from frame losses by playing contiguous frames in quick succession, creating a perception of \mquote{fast} movement. Both these temporal artefacts are grouped into freeze. 
This happens mainly due to network loss (where some frames or blocks lost) or delay (where the frames are dropped because the frames are decoded too late to display). 

\noindent \textbf{Blurriness} appears when encoder uses high quantization parameters (QP) during transform coding. Typically, servers use adaptive encoding based on network conditions. In adaptive encoding, server attempts to adjust QP to minimize bitrate in poor network conditions, which degrades the quality of encoded frame. High QP reduces the magnitude of high frequency DCT coefficients almost down to zero, consequently losing information and making it impossible to extract original DCT coefficients at the time of de-quantization.  Loss of high-frequency information results in blurriness. 

\noindent \textbf{Blocking:} Most of the current coding standards (such as H.26x and VPx) are block based, where each image is divided into blocks (from 4x4 to 32x32 and recent 64x64 block in H.265). The blocking metric captures the loss of these blocks due to high network-packet loss. The decoder can introduce visual impairments at the block boundaries or place random blocks in place of original blocks, if the presentation timestamp elapses, which creates bad experience. 

\noindent \textbf{Call startup delay} is the duration of call setup from the time the caller initiates the call until the callee receives it. 
We observe an 11 s worst case and 7 s median delay when there is 20\% network packet loss, whereas we obtain a median 3 s delay in best network conditions.
Although call startup delay does measure user experience, it can only provide information at the beginning of the call, and not during the call.

\noindent \textbf{Audio and video synchronization:} The relationship between audio and video is important in case of interactive video telephony. For example, most of the telephony applications allow audio playback while the video is being frozen. The QoE can be formulated as a complex model in terms of audio and video metrics. However, most of these applications separate the audio and video streams because of the advances in adaptive bitrate algorithms. Also, considering the audio traffic is very small, we focus only on the video quality of experience excluding audio streams. We do not quantify the audio experience as there is an extensive work focused on this in the past \cite{chen2006quantifying}\cite{yu2014can}.

In this work, we focus on freeze and blurriness artefacts. In our extensive experiments over Skype, Hangouts, FaceTime, we do not observe any blocking artefacts, hence we omit this metric for video telephony. We notice video freezes (temporal artefact) and blurring (spatial artefact). For video freezes, we  explore multiple metrics that capture freeze ratio for whole clip, number of freezes per clip and duration of freezes. We do not consider startup delay, as it is an one-time metric.

\begin{figure}[t]
  \centering
  \includegraphics[width=\linewidth]{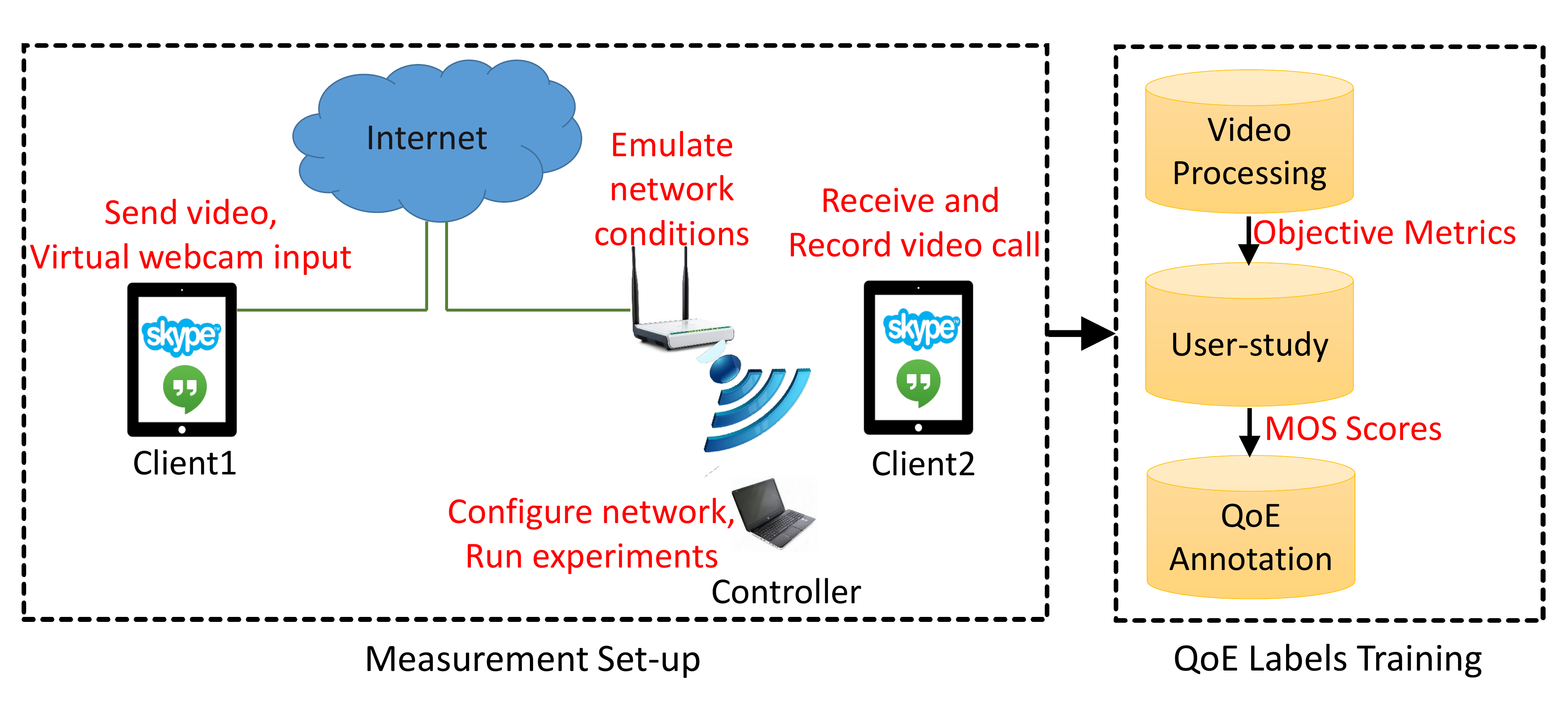}
     \vspace{-0.3in}
    \caption{Measurement set-up and system architecture. \textit{Left:} Video telephony measurements and recording video calls. \textit{Right:} Our framework processing recorded videos offline and extracting objective metrics to predict QoE labels.}
   \vspace{-0.2in}
  \label{fig:set-up}
\end{figure}
\begin{figure}[t]
  \centering
  \includegraphics[width=\linewidth]{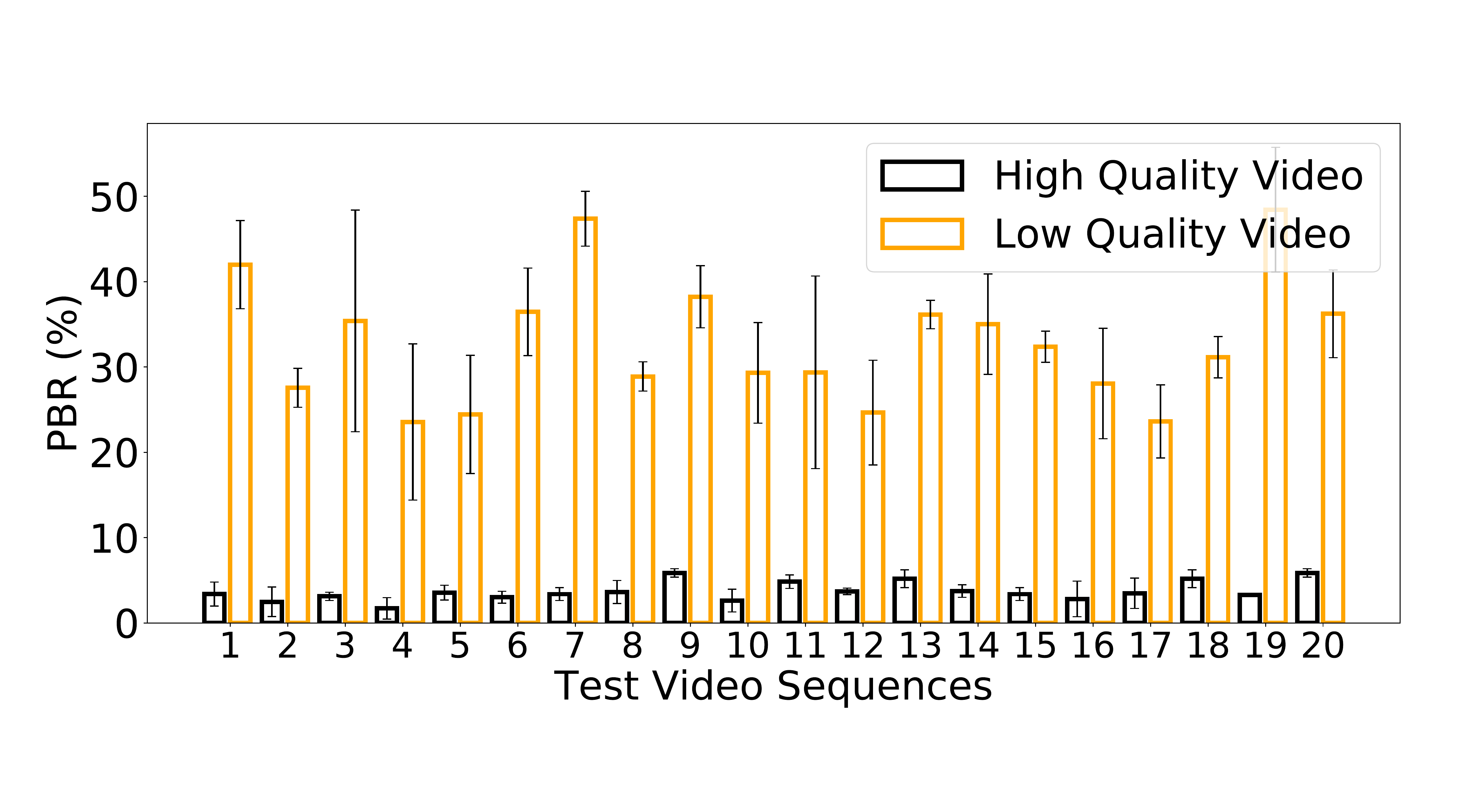}
   \vspace{-0.5in}
  \caption{Blur detection using PBR.}
\vspace{-2em}
  \label{fig:ourblurdetection}
\end{figure}

\section{Design for Scalable QoE Annotation} \label{label:design}
In this section, we first describe our framework and measurement methodology. 
We then explain our QoE metrics and evaluate them across different applications and devices. We validate that our metrics capture video artefacts caused by diverse network conditions.
\subsection{Measurement Methodology}

\noindent \textbf{Set-up:} 
In Figure~\ref{fig:set-up}, we show our measurement set-up and QoE labeling framework. 
Since video telephony is an interactive application, it requires at least two participating network end-points (clients). In our set-up, we use a mobile device on our local enterprise \wifi network (Client 1) and an Amazon-provided instance as the second end-point (Client 2). Both clients can run Skype or Hangouts\footnote{For FaceTime, we use 2 local clients (iPhone/iPad and MacBook Air) on different sub-networks, as creating virtual iOS and web-cam is challenging.}. 
When the video call is placed from Client 1 to Client 2, Client 2 runs a virtual web-cam that inputs a video file in telephony application instead of camera feed; at Client 1, the video is received during the call.
At the Amazon Client 2, we use \textit{manycam} \cite{mancam}, a virtual web-cam tool. 
This tool can be used with multiple video applications in parallel for automation. 
In our LAN, a controller sits  to emulate diverse network conditions and to run experiments on the connected mobile device  through Android debug bridge (ADB) interface (via USB or \wifi connection). 
To automate the video call process, we use AndroidViewClient (AVC) library \cite{awc}.
Once the received call is accepted, we start the screen recording of the video call session.   The videos are recorded using Az screen recorder \cite{azscreen}. 
The recorded videos are sent to video processing module to calculate the objective QoE metrics. 
We use Ffmpeg \cite{ffmpeg} tool to extract our QoE metrics. Ffmpeg is a video framework with large collection of coding libraries. To validate our metrics and translate them into user experience, the videos are then shown to users to rate their experience. 
Finally, we retrieve MOS from all users and map our QoE metrics to MOS. We delve into the details of our user  study and modeling in Section~\ref{label:results}.


\begin{figure*}
    \centering
    \vspace*{-1em}
    \includegraphics[trim={2.5cm 24.5cm 2.5cm 0},clip,width=0.8\linewidth]{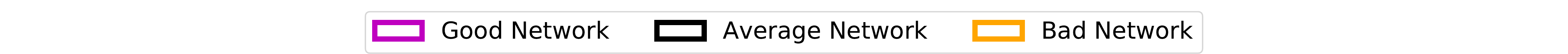}
  \subfigure[Content and Motion]{\hspace*{-1em}\includegraphics[width=0.33\textwidth]{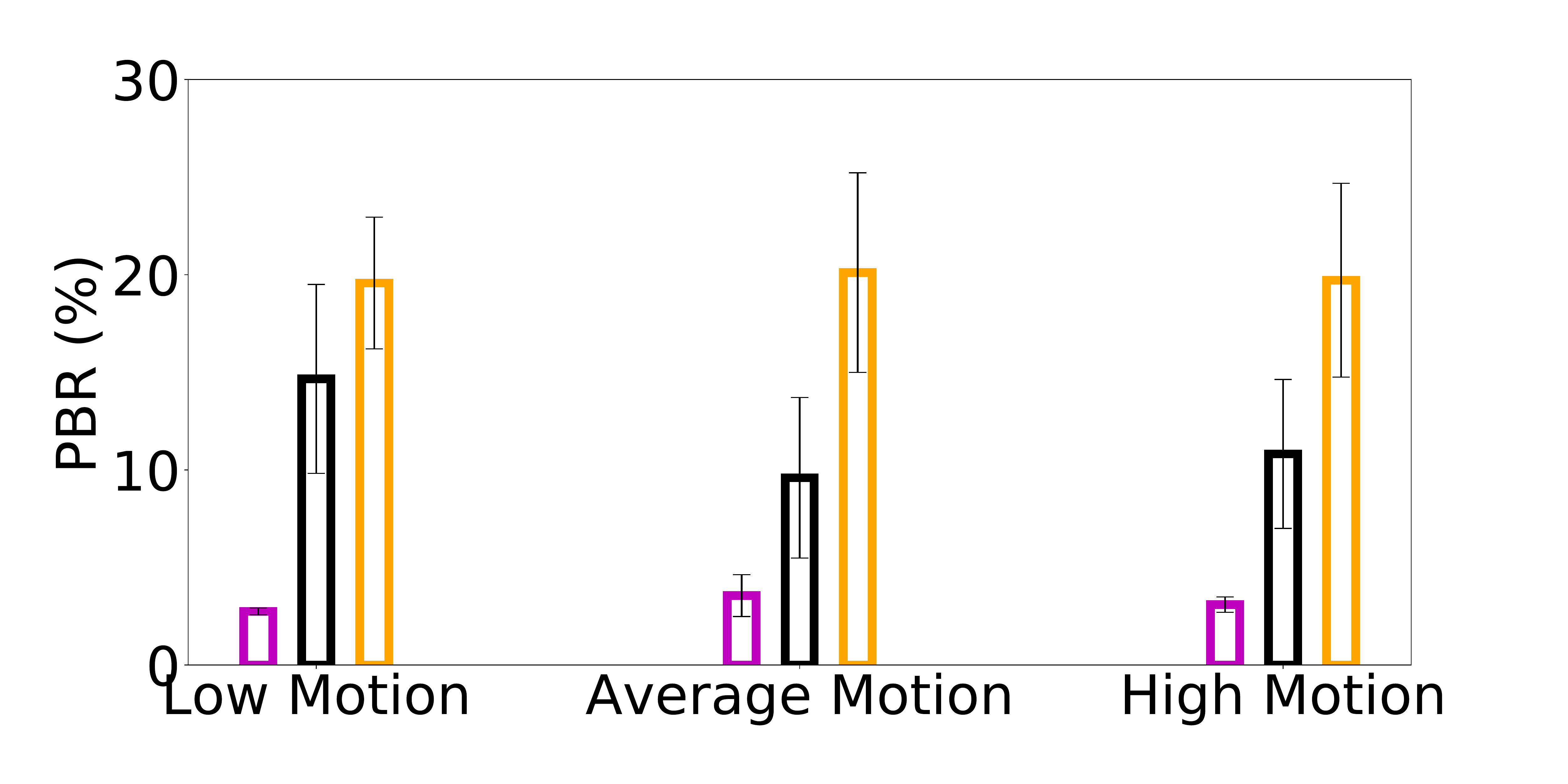}\label{fig:cont-pbr} \vspace*{-3em}}~
  \subfigure[Applications]{\hspace*{-1em}\includegraphics[width=0.33\textwidth]{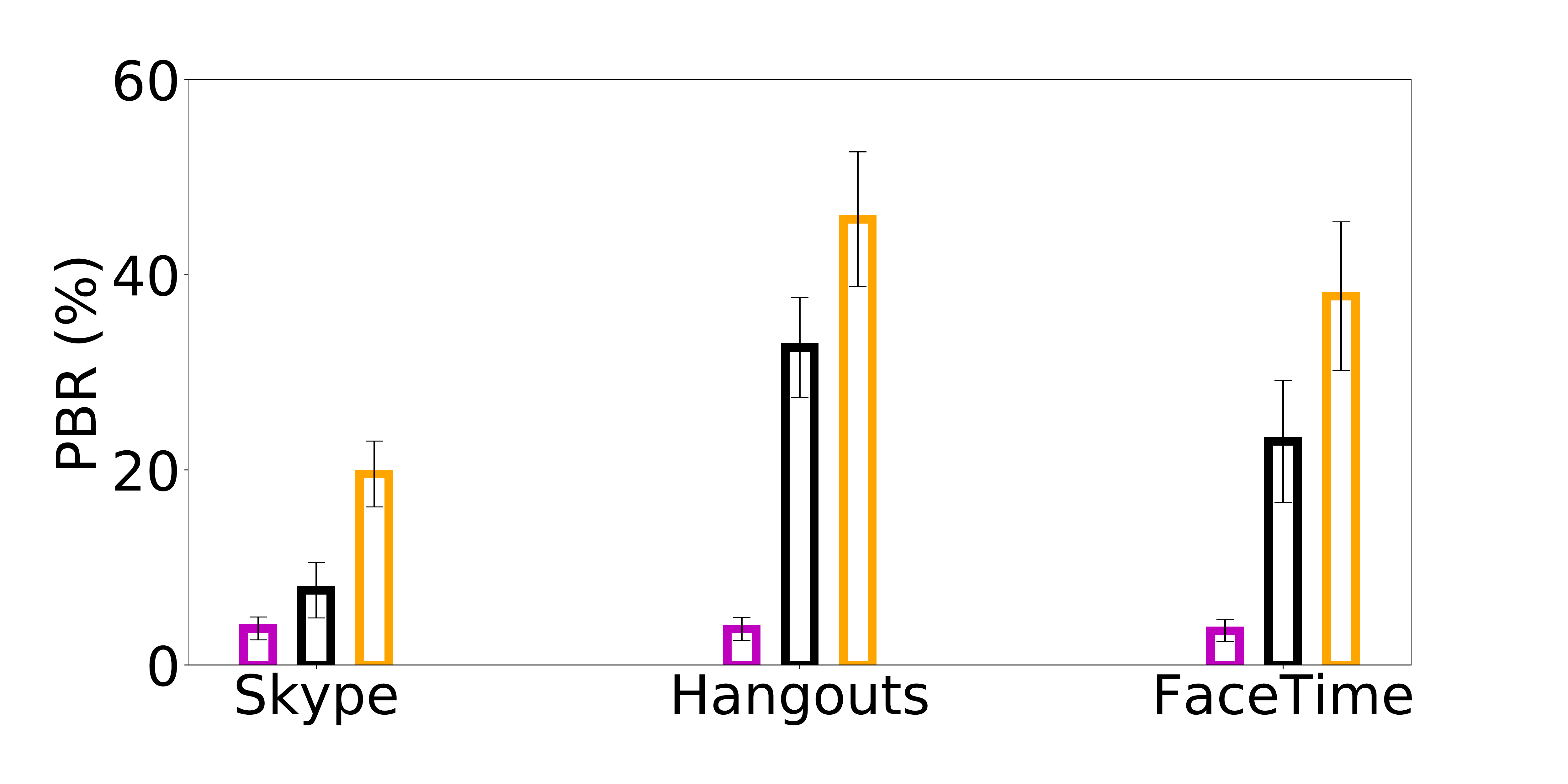}\label{fig:app-pbr} \vspace*{-3em}}~
   \subfigure[Devices] {\hspace*{-1em} \includegraphics[width=0.33\textwidth]{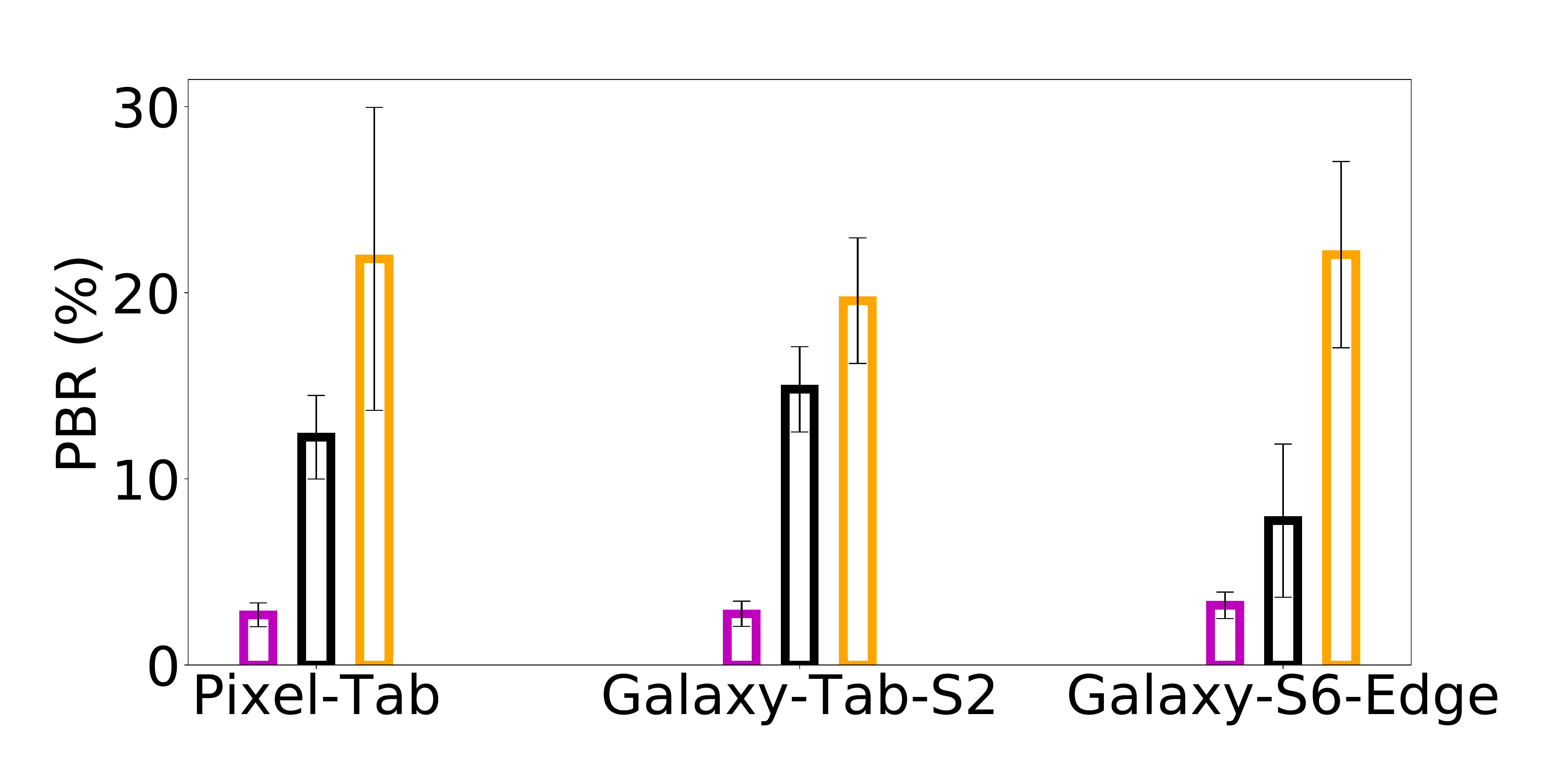}\label{fig:dev-pbr} \vspace*{-3em}}
     \subfigure[Content and Motion]{\hspace*{-1em}\includegraphics[width=0.33\textwidth]{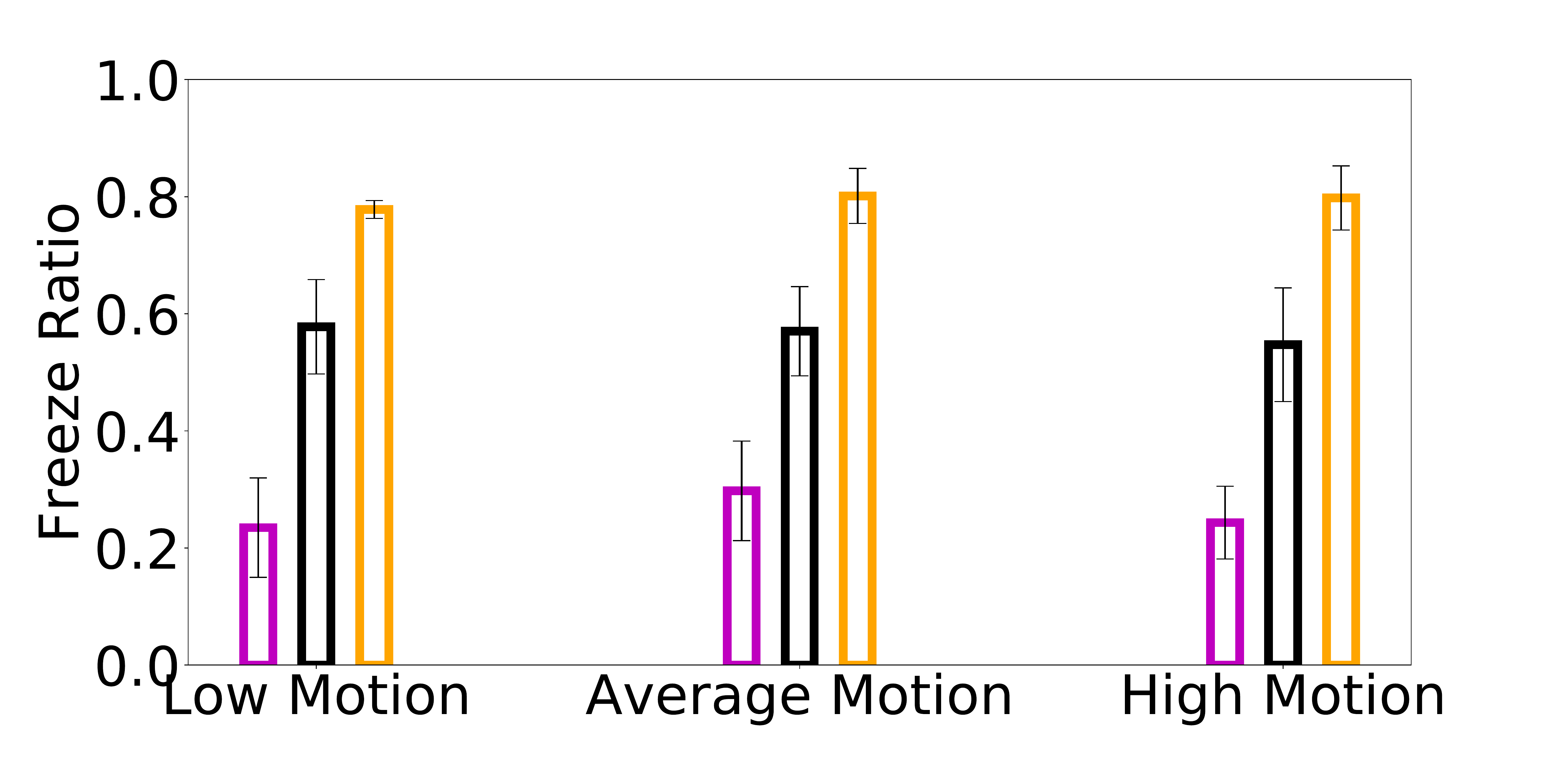}\label{fig:cont-st}}~
     \subfigure[Applications]{\hspace*{-1em}\includegraphics[width=0.33\textwidth]{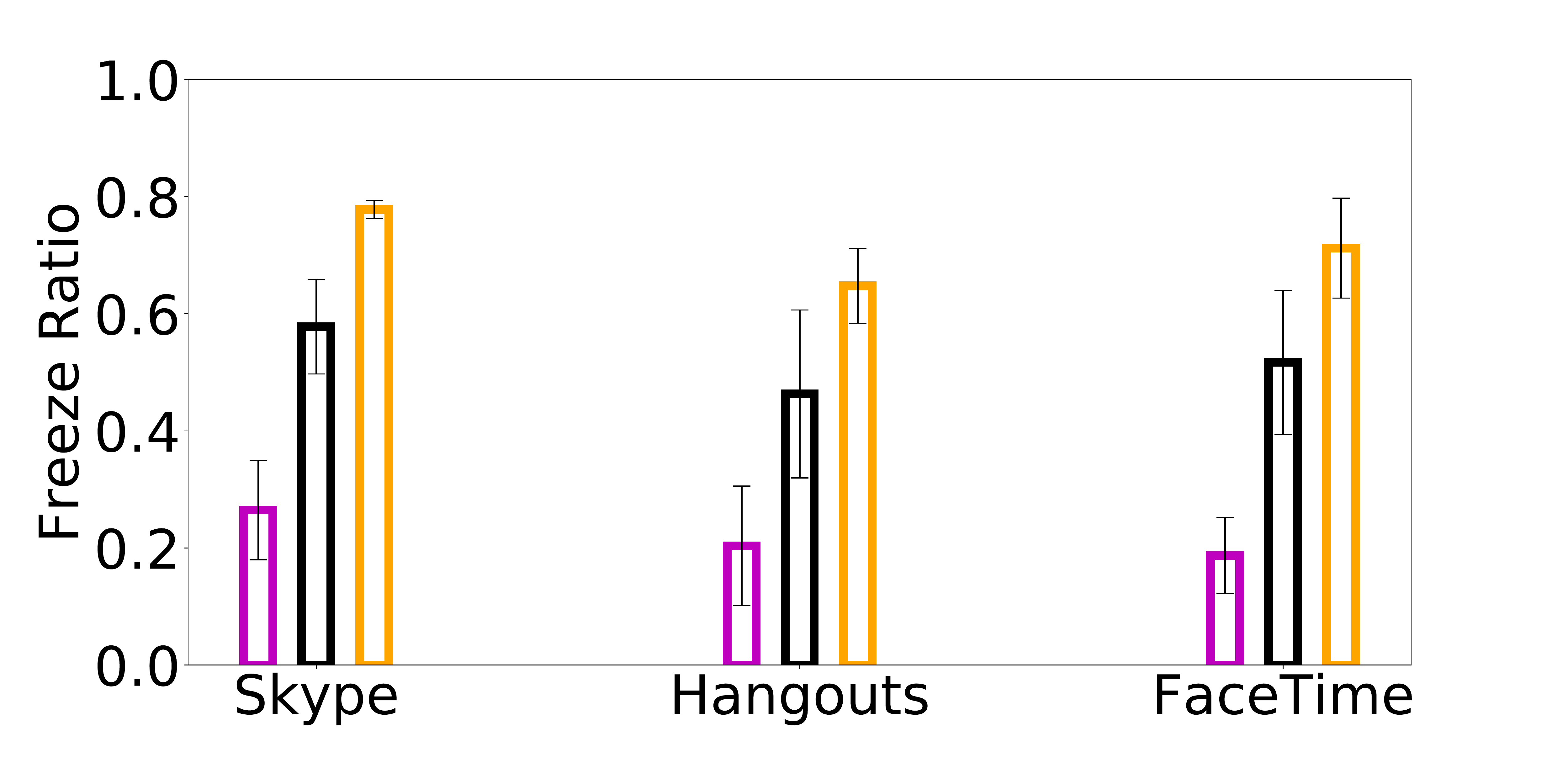}\label{fig:app-st}}~
     \subfigure[Devices] {\hspace*{-1em} \includegraphics[width=0.33\textwidth]{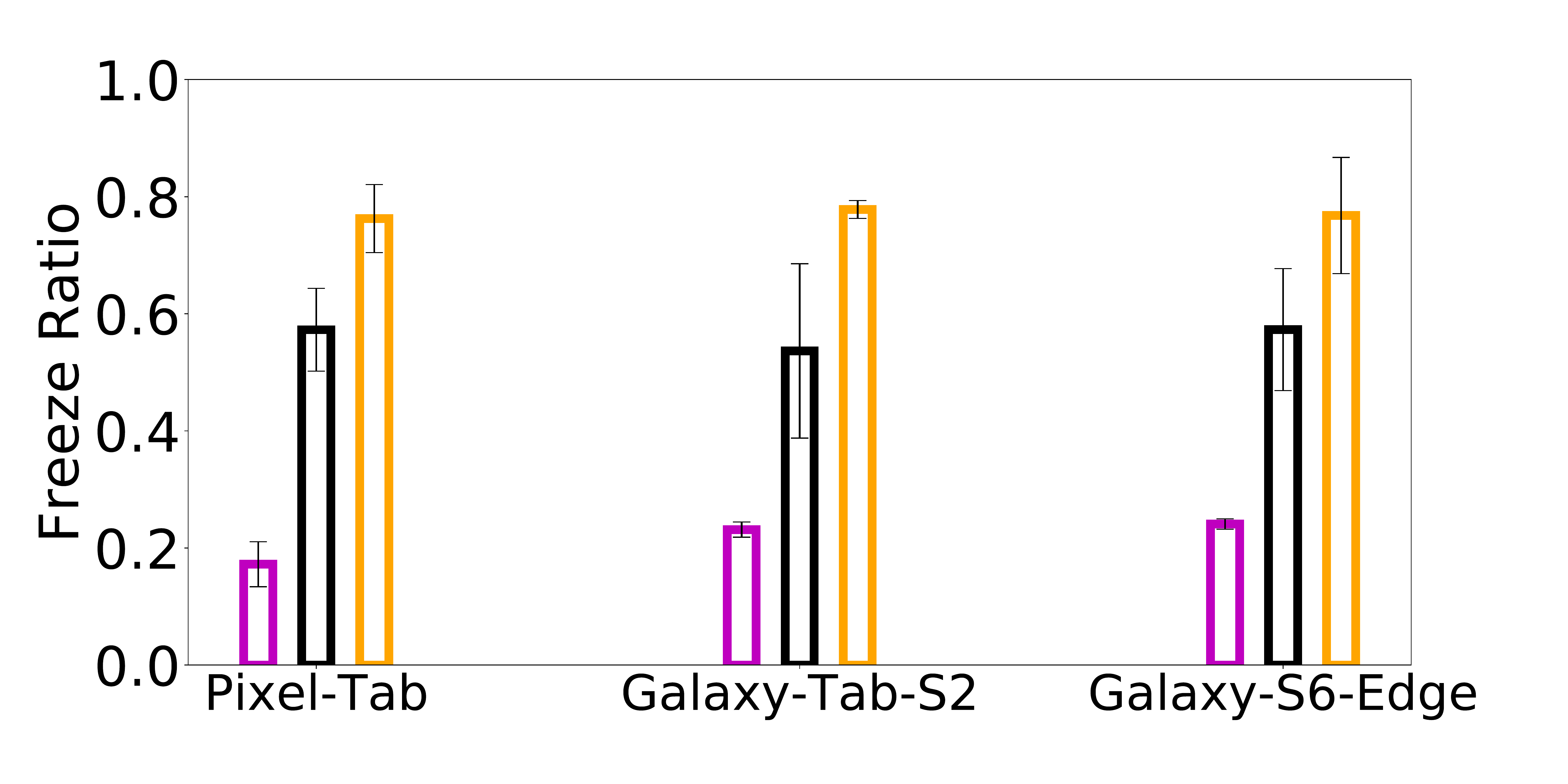} \label{fig:dev-st}}
    \caption{PBR and Freeze Ratio for different videos with respect to content, application and device settings. Results for diverse network conditions are shown (from bad, average, good).}
     \label{fig:measurements}
     \vspace*{-1em}
\end{figure*}

\noindent \textbf{Network conditions:} As we experiment real-time with interactive applications (Skype, Hangouts, FaceTime), we need to emulate real user experience under any condition and to obtain samples with MOS values of all levels (1--5). Hence, we conduct tests with ideal network conditions and with throttled network. To create average and bad network conditions, we use Linux utility \texttt{tc} and we introduce hundreds of ms of delay, packet loss (10--20\% ensuring that the call is setup), or low bandwidth (2--10 Mbps). Note that the network paramters in general can be varying considerably. But, we by design choose these non-varying settings to have a controlled environment to get reliable ground-truth about the quality of experiments. We need a similar loss environment for 10-30 second clips in order to be able to accurately use the ground-truth from the users.  

\noindent \textbf{Test video sequences:} We use the same 20 test video sequences as in Section \ref{MOTIVATION}. 
We collect these videos  from Xiph media \cite{xiph2008org} and YouTube. We select 20 representative videos that contain different types of motion and content.
All videos are downloaded in Full HD (1920x1280) resolution.

\subsection{Video QoE Metrics}

\noindent \noindent \textbf{PBR:} 
Video bitrate has been considered a standard metric for perceptual quality of video \cite{zhang2012profiling, chakraborty2016exbox}.
We compute the bitrate of recorded videos as a QoE metric. Typically, the bitrate is lower for low resolution video and higher for high resolution video and depends on the level of video compression.
Therefore, we capture video blur with \texttt{encoded bitrate} by compressing the recorded video. 
We employ the observation that the blurrier the video is, the higher compression efficiency is. 
However, we find that the bitrate metric is sensitive to motion in the video, because the block movement for high motion videos is very high and it is low for low motion videos. This results in different encoding bitrates. 
As described in Section \ref{label:background}, low motion videos take advantage of inter-frame prediction, which makes the encoded bitrate motion-sensitive. 
Therefore, we use intra-coded bitrate by \emph{disabling the inter frame prediction while compressing video.}
We experiment with different encoding parameters like quantization parameter (QP) and de-blocking filter techniques and we choose high QP value (30) while coding, to get large bitrate difference when encoding high- and low-quality videos. 
To achieve robustness to video content, we compute the \emph{relative change between the recorded bitrate and  the intra-coded bitrate of the compressed video.} We define this change as the perceptual bitrate (PBR), as it only captures quality of the image. Since we calculate PBR on the recorded video, we do not need a reference video, hence PBR is a no-reference metric.

We validate the PBR metric with respect to the same 20 videos used to compute blur using DCT coefficients in Section \ref{MOTIVATION}.
We create two sets of videos with high and low quality and record them during a Skype call under best network conditions.
This experiment aims to validate our blur capturing metric, hence we avoid video freezes by sending low quality videos over best network conditions.
Fig. \ref{fig:ourblurdetection} shows PBR for both low and high quality videos.
We observe a PBR larger than 20\% for low quality videos, whereas PBR is smaller than 5\%  for high quality videos.
Unlike blur detection using previous methods, we see a clean separation of average PBR between low and high quality videos. Therefore, PBR can be used to detect blurriness in videos without any ambiguity.


\noindent \textbf{Freeze Ratio:} As described in Section \ref{label:background}, freeze ratio is another important QoE artefact. 
Freeze ratio is the number of repeated frames over total frames in a given window, i.e., it denotes the amount of time the video is paused because of network disturbance. 
We use Ffmpeg's \textit{mpdecimate} \cite{ffmpeg} filter in calculating freeze ratio. 
The \textit{mpdecimate} algorithm works as follows: The filter divides the current and previous frame into 8x8 pixels blocks  and computes sum of absolute differences (SAD) for each block. 
A set of thresholds (\textit{hi, lo and frac}) are used to determine if the frames are duplicate. 
The thresholds \textit{hi} and \textit{lo} represent number of 8x8 pixel differences, so a threshold of 64 means 1 unit of difference for every pixel. A
frame is considered to be duplicate frame if none of the 8x8 blocks yields SAD greater than a threshold of \textit{hi}, and if no more than \textit{frac} blocks have changed by more than \textit{lo} threshold value. 
We experiment with several other error methods such as MSE and MAD, but we notice similar results among all three, thus we choose SAD for minimal computation overhead. 
We use threshold values of 64*12 for \textit{hi}, 64*5 for \textit{lo} and 0.1 for \textit{frac} for all our experiments. 
The metric does not work if the entire video is having a still image (for instance a black screen throughout the video). 
However, we think that it is a reasonable assumption that most of video telephony applications do not generate such video content.

In addition to freeze ratio, we also compute length and number freezes in video. We define a freeze if the video is stalled for more than one second. We observe that users do not perceive the stall when the length of freeze is very short or if there are very few such short freezes in the video. Therefore, we employ \texttt{freeze ratio, length and number of freezes} metrics to capture the temporal artefacts of QoE.

\subsection{Micro-Benchmarking of Video Artefacts}
In this section, we show the scalability of our metrics through measurements across different applications and devices. As described above, we use a total of 4 metrics: PBR, freeze ratio, length and number of freezes in the video. For brevity, we show measurements for only PBR and freeze ratio. We find similar trends with other metrics as well. We measure these metrics with 6 different videos from the 20 videos described in Section \ref{MOTIVATION}, that cover high video-motion and content diversity. These experiments are run on Skype and Samsung Galaxy Tab S2 unless otherwise specified. All videos are recorded under Full HD (1920x1280) resolution with 60 Fps. Each experiment consists of 20 minutes video call with a total of 18 hours of video recordings. The videos are recorded under different network conditions to obtain different video quality  levels. We use the following network conditions: good case (0\% loss, 0 latency and 100 Mbps bandwidth), average case (5\% loss, 100 ms and 1 Mbps bandwidth) and bad case (20\% loss, 200 ms, and 512 Kbps bandwidth). Fig. \ref{fig:measurements} shows PBR and freeze ratio across different applications and devices. Our observations are the following.

\begin{figure*}[t]
  \centering
  \includegraphics[width=\linewidth]{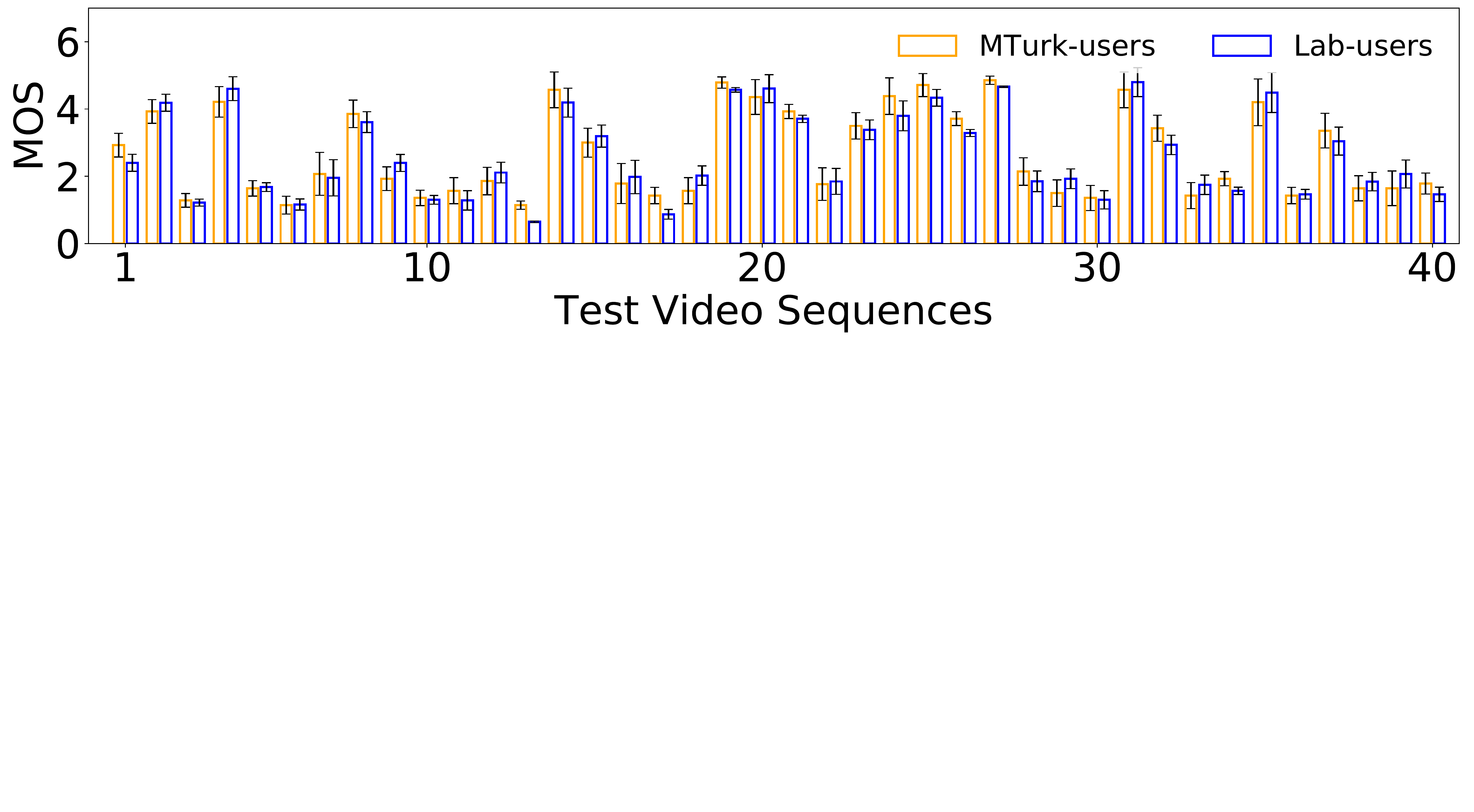}
  \vspace{-6cm}
  \caption{MOS and standard deviation of QoE scores from 30 lab and 30 online users. Although the scores vary across the users both in lab and online, the distribution of MOS is similar in two cases.}
  \vspace{-0.15in}
  \label{fig:mos}
\end{figure*}

\noindent \textbf{Video Motion and Content Diversity:} 
The average PBR for best network conditions is always smaller than 5\%. PBR is larger than 20\% under bad settings (see In Fig.~\ref{fig:cont-pbr}). 
The average freeze ratio for best network conditions is smaller than 0.3 and for bad conditions, it is larger than 0.8 (Fig.~\ref{fig:cont-st}). 
We observe that both PBR and freeze ratio are highly impacted by network conditions and follow same trend with network quality irrespective of content and motion in the video.
Therefore, the videos can be labeled accurately because of clear separation between network dynamics.

\noindent \textbf{Application Diversity:} 
In Fig.~\ref{fig:app-pbr},~\ref{fig:app-st}, we present PBR and freeze ratio for three applications: Skype, FaceTime and Hangouts.
For Skype, we observe an average PBR of less than 5\% with good network and 20\% with bad network. 
Whereas, both FaceTime and Hangouts reach more than 40\% PBR under bad network. 
This discrepancy is due to application logic: Skype does not compromise quality of video, hence low PBR. But, this causes Skype video calls to be stalled more frequently compared to other applications under bad network scenarios. Whereas,  bitrate adaptation of FaceTime and Hangouts sacrifices quality to provide temporally smooth video. 
Moreover, quality and bitrate are highly impacted by the underlying video codec the application uses. In our experiments, Skype uses H.264 whereas Hangouts uses VP8. We observe that using different video codecs and adaptive bitrate (ABR) algorithms impacts video quality during call.

The freeze ratio for Skype is similar to FaceTime and Hangouts. However, as network conditions worsen, Skype yields more freezes compared to other applications. Despite the additional freezes, the average freeze-ratio for Skype is only 0.1 larger than other applications. This is due to the longer freezes of FaceTime and Hangouts than of Skype.

\noindent \textbf{Device Diversity:} We aim to devise a model that is independent of the device that runs video telephony. Our metrics should capture similar video QoE on different devices given that screen recorder and network conditions are the same. To this end, we evaluate our metrics on three devices: Samsung Galaxy Tab S2, Samsung Galaxy Edge S6 and Google Pixel Tab. 
In Fig.~\ref{fig:dev-st},~\ref{fig:dev-pbr}, we observe that both PBR and freeze ratio show same distribution under different network settings on all devices. Hence, our metrics are device independent.

\section{From Video Artefacts to Q\lowercase{o}E} \label{label:results}
We validate our QoE metrics presented in Section~\ref{label:design} with subjective measurements. We seek to ensure that our metrics accurately capture  video telephony QoE artefacts. Video artefacts vary across applications and network conditions. Additionally, they can appear together (e.g., when we have both blurriness and stutter) or independently. Intermingled video artefacts result in diverse QoE levels (from bad to excellent), standardized as MOS by ITU-T~\cite{series2012methodology}\footnote{MOS takes values from 1 to 5, with 5 being excellent.}. We carry out a user study and obtain a MOS for each video.

\subsection{Data Preparation}
As described in Section \ref{label:design}, the dataset is prepared from recorded video calls. We use 20 videos described in Section \ref{MOTIVATION}, for video calls on Skype, FaceTime and Hangouts. For each video and application, we repeat  experiments on  multiple devices. The video calls are recorded under different network conditions to produce good, average and bad quality videos. We record video calls for Skype, Hangouts on Samsung Galaxy (SG) S6 edge, Google Pixel Tab, SG S2 Tab and for FaceTime on iPad Pro (model A1674) and iPhone 8 Plus. We post-process these videos into 30 seconds  video clips and evaluate QoE for each clip. We host a web server with these video clips and ask the users to rate their experience after watching each sample. 

\subsection{User-Study Setup}
We conduct the user-study in two phases: 1) in two labs, 2) online using a crowd sourced platform. The lab user study is conducted at a university campus and an enterprise lab. The online user study is conducted on Amazon Mechanical Turk platform \cite{turk-amazon}. Online users are from the United States and are in age range 20--40. We collect results from 60 lab users and more than 150 online users\footnote{Our conducted user-studies are approved by the Institutional Review Board (IRB) of our institution. Evidence to be provided upon paper acceptance.}. We faced several challenges  while deploying a user-study in the wild, such as ensuring that 1) users watch the whole video clip without fast forwarding and then rate their experience, 2) users can rate the video if and only if they watch the video, 3) users watch on screens of similar size  to avoid huge QoE variance i.e., small screen users do not perceive poor quality compared to large screen users. We address these challenges by carefully designing the user-study website and avoiding noisy samples. First, we disable the video controls in the web page so that users cannot advance the video. Second, we introduce random attention pop-up buttons while watching the video and ask users to click on the attention button to confirm that they are watching the video. Note that the pop-up button is visibile for very few seconds below the video playback, which does not have a significant influence while rating the video (See \S \ref{resultstudy} for variance results). A user cannot move to next video until the current video is rated. Finally, we restrict users from taking the study on mobile devices, to avoid variance due to small screens. The user-study web page can be run on Chrome, Firefox and Safari. Once the user completes rating all the videos, results are pushed to our server. Then, users are granted a server-generated code for processing payments in MTurk. Amazon MTurk restricts users from taking again the same study, thus our users are unique.

\subsection{Lab  vs.~MTurk Users} \label{resultstudy}
We compare the MOS across lab and online users for the same set of 40 video clips and we find similar distributions across 80 clips. Fig. \ref{fig:mos} shows the MOS bar plots of user ratings. The lower the rating is, the lower the QoE is. For each video sequence, we plot the average MOS across users with error bars for standard deviation. We observe that lab and online users exhibit similar distributions of MOS. The standard deviation of MOS between lab and online users is smaller than 1 MOS for 96\% of videos. 
In both studies, we observe lot of variance for first few videos because the users were not shown original videos. The users learn the score for each video after watching few videos and rate them correctly from then on. So we remove these initial set of videos which have high standard deviation from our data analysis. 
We find that 4\% of these video clips have more than 1 MOS standard deviation and we remove them from our analysis as outliers. Such videos have a  large portion of uniform background, hence their content challenges classification between average and good MOS.

\subsection{Modeling Our Metrics to MOS} \label{label:model}

We now present our MOS prediction model that corroborates that our metrics accurately capture QoE artefacts and users' MOS. We build a model to map our objective metrics to subjective evaluations (MOS from user-study). Typically network administrators need to estimate a subjective evaluation such as MOS. However, QoE assignment in 5 classes can be cumbersome and may give little information on network quality. Hence, we map MOS scores to 3 classes (i.e., bad, average, good) that can be used by LTE or \wifi deployments to enhance resource allocation. We first explain our modeling methodology from our metrics to MOS, and then describe how to translate MOS into 3-class QoE.

Typically, objective QoE metrics are mapped to MOS scores using non-linear regression \cite{cui2008image}. 
Our regression model employs the average MOS from 15 users as ground-truth per video clip and it is based on ensemble methods \cite{ensemblemethods2008}. 
Taking the ensemble of models, we combine predictions of base estimators to improve generalizability and robustness over a single model.
Moreover, a single model is always vulnerable to over-fitting and is complicated, which can be avoided in the form of ensemble of weak models.
The ensemble is usually produced by two methods: \textit{averaging} or \textit{boosting}. 
We employ boosting method, in particular \texttt{AdaBoost}\cite{freund1997decision}, where the base estimators are built sequentially and one tries to reduce the bias of the combined estimator, thereby combining several weaker models and producing an accurate model. 

\texttt{AdaBoost} algorithm fits a sequence of weak models (for example, small decision trees in our framework nearly equivalent to random guessing) on different versions of data. 
The predictions from each weak model are combined to produce a strong prediction with a weighted sum scheme. 
At each iteration, called boosting iteration, a set of weights ($w_1, w_2, …, w_N$, where $N$ is number of samples) are applied to training data.
The weights are initialized with $w_i = 1/N$, in order to train a weak model on the original data.
In the following iterations, the weights of training samples are individually modified and model is applied again on the re-weighted data.
At any given boosting stage, the weights are changed depending on prediction at the previous step.
The weights are proportionately increased for incorrectly predicted training samples, while the weights are decreased for those samples that were correctly predicted.
The samples which are difficult to predict become more important as the number of iterations increases.
The next-iteration weak model then focuses on the incorrectly predicted samples in the previous step.
Details of the Adaboost algorithm are given in \cite{freund1997decision}\cite{drucker1997improving}.

Moving from 5 MOS scores to 3 classes, one has to estimate two thresholds $m_1, m_2$ in MOS (bad label: MOS  $ < m_1,$ average: $m_1 \le$ MOS $< m_2,$ good: MOS $\ge m_2$).
To this end, we first train the regression model and then search the MOS scores space with a sliding window of 0.05 for the two thresholds. 
We iterate over all such possible thresholds to maximize the accuracy of the trained model. 
We compute the true labels and prediction labels from the test MOS scores and predicted scores respectively using the corresponding thresholds in each iteration. 
We then select the thresholds which give highest accuracy from prediction labels.
Therefore, our framework is divided into two phases: predicting MOS scores and labeling the scores with optimal thresholds.

\begin{table}[t]
    \centering
    \caption{Models performance on Skype data}
    \label{class-skype}
    \begin{tabular}{|l|l|l|l|l|}
        \hline
        \textbf{Model} & \textbf{Precision (\%)} & \textbf{Accuracy (\%)} & \textbf{Recall (\%)} & \textbf{MSE} \\ \hline
        SVR      & 89.33                      & 89.33                   & 89.33                     & 0.36                 \\ \hline
        MLP      & 90.77                      & 90.77                   & 89.62                     & 0.36                  \\ \hline
        KNN      & 82.91                      & 81.67                   & 81.67                     & 0.63                  \\ \hline
        RF       & 89.72                      & 89.34                   & 89.34                     & 0.41                  \\ \hline
        ADT       & 92.21                      & 92.00                   & 92.00                     &0.29                   \\ \hline 
    \end{tabular}
\end{table}

\begin{table}[t]
    \centering
    \caption{Models performance on FaceTime data}
    \label{class-facetime}
    \begin{tabular}{|l|l|l|l|l|}
        \hline
        \textbf{Model} & \textbf{Precision (\%)} & \textbf{Accuracy (\%)} & \textbf{Recall (\%)} & \textbf{MSE} \\ \hline
        SVR      & 88.13                   & 86.00                  & 86.00                &  0.32                 \\ \hline
        MLP      & 90.00                   & 89.77                  & 89.30                &  0.28                 \\ \hline
        KNN      & 74.98                   & 72.00                  & 72.30                &  0.48                 \\ \hline 
        RF       & 90.37                   & 90.00                  & 90.00                &  0.28                 \\ \hline 
        ADT       & 90.28                   & 90.00                  & 90.00                & 0.26                  \\ \hline 
    \end{tabular}
\end{table}

\begin{table}[t]
    \centering
    \caption{Models performance on Hangouts data}
    \label{class-hangouts}
    \begin{tabular}{|l|l|l|l|l|}
        \hline
        \textbf{Model} & \textbf{Precision (\%)} & \textbf{Accuracy (\%)} & \textbf{Recall (\%)} & \textbf{MSE} \\ \hline
        SVR            & 91.36                        & 90.67                       & 90.67                     & 0.44                  \\ \hline
        MLP            & 89.34                        & 88.48                       & 88.48                     & 0.45                  \\ \hline
        KNN       & 87.02                        & 86.67                       & 86.67                     & 0.49                  \\ \hline 
        RF       & 91.22                        & 90.67                       & 90.67                     &  0.43                 \\ \hline 
        ADT       & 93.75                        & 93.33                       & 93.33                     & 0.38                  \\ \hline 
    \end{tabular}
\end{table}

\begin{table}[t]
\centering
\caption{Model performance across devices for Skype}
\label{label:skype-devices}
    \resizebox{\columnwidth}{!}{
    \begin{tabular}{|c|c|c|c|c|}
\hline
        \multicolumn{2}{|c|}{\textbf{Devices}}                & \multicolumn{3}{c|}{\textbf{Model Performance}}             \\ \hline
    \multicolumn{1}{|c|}{Training} & Testing     & \multicolumn{1}{c|}{Precision (\%)} & {Accuracy (\%)} & {Recall (\%)} \\ \hline
\multirow{3}{*}{SG-S6 Phone}   & SG-S6 Phone & 89.90  & 88.57 & 88.57       \\ \cline{2-5} 
                               & SG-S2 Tab   & 88.41  & 88.00 & 88.00       \\ \cline{2-5} 
                               & Pixel Tab   & 88.52  & 87.62 & 87.62       \\ \hline
\multirow{3}{*}{SG-S2 Tab}     & SG-S6 Phone & 83.73  & 84.43 & 83.00       \\ \cline{2-5} 
                               & SG-S2 Tab   & 93.04  & 91.43 & 91.43       \\ \cline{2-5} 
                               & Pixel Tab   & 82.69  & 82.86 & 82.86       \\ \hline
\multirow{3}{*}{Pixel Tab}     & SG-S6 Phone & 84.43  & 84.00 & 84.00       \\ \cline{2-5} 
                               & SG-S2 Tab   & 84.40  & 86.49 & 86.49       \\ \cline{2-5} 
                               & Pixel Tab   & 88.20  & 94.40 & 94.00       \\ \hline
\end{tabular}
}
\end{table}

\begin{table}[t]
\centering
\caption{Model performance across devices for FaceTime}
\label{label:facetime-devices}
    \resizebox{\columnwidth}{!}{
\begin{tabular}{|c|c|c|c|c|}
\hline
    \multicolumn{2}{|c|}{\textbf{Devices}}                & \multicolumn{3}{c|}{\textbf{Model Performance}}             \\ \hline
    \multicolumn{1}{|c|}{Training} & Testing     & \multicolumn{1}{c|}{Precision (\%)} & {Accuracy (\%)} & {Recall (\%)} \\ \hline
\multirow{3}{*}{iPad Pro}   & iPad Pro & 93.60  & 92.00 &   92.00     \\ \cline{2-5} 
                               & iPhone 8 Plus & 90.18  & 89.93 & 89.93       \\ \hline
\multirow{3}{*}{iPhone 8 Plus}     & iPad  Pro  & 88.34  & 88.34  & 88.34       \\ \cline{2-5}  
                               & iPhone 8 Plus  & 92.50  & 92.00 & 92.00       \\ \hline
\end{tabular}
}
\end{table}

\begin{table}[t]
\centering
\caption{Model performance across devices for Hangouts}
\label{label:hangouts-devices}
    \resizebox{\columnwidth}{!}{
\begin{tabular}{|c|c|c|c|c|}
\hline
    \multicolumn{2}{|c|}{\textbf{Devices}}                & \multicolumn{3}{c|}{\textbf{Model Performance}}             \\ \hline
    \multicolumn{1}{|c|}{Training} & Testing     & \multicolumn{1}{c|}{Precision (\%)} & {Accuracy (\%)} & {Recall (\%)} \\ \hline
\multirow{3}{*}{SG-S6 Phone}   & SG-S6 Phone & 90.03  & 90.00 & 90.00       \\ \cline{2-5} 
                               & SG-S2 Tab   & 84.26  & 84.77 & 84.77       \\ \cline{2-5} 
                               & Pixel Tab   & 82.86  & 82.00 & 82.00       \\ \hline
\multirow{3}{*}{SG-S2 Tab}     & SG-S6 Phone & 89.61  & 89.21 & 89.21       \\ \cline{2-5} 
                               & SG-S2 Tab   & 91.76  & 90.00 & 90.00       \\ \cline{2-5} 
                               & Pixel Tab   & 85.41  & 84.77 & 84.77       \\ \hline
\multirow{3}{*}{Pixel Tab}     & SG-S6 Phone & 86.79  & 86.00 & 86.00       \\ \cline{2-5} 
                               & SG-S2 Tab   & 85.05  & 85.00 & 85.00       \\ \cline{2-5} 
                               & Pixel Tab   & 86.17  & 86.67 & 86.67       \\ \hline
\end{tabular}
}
\end{table}

We first evaluate our metrics by fitting five most common regressors: Support Vector Regressor (SVR), Random Forests (RF), Multi-Layer Perceptron (MLP), K-Nearest Neighbor (KNN) and Adaboosted Decision Tree Regressors (ADT). Each model is evaluated under 10 fold cross-validation. We perform a fine grid search to tune the hyper-parameters for all the models. We select the best parameters from the grid search and use the best estimator for the rest of the evaluations. This is repeated for Skype, FaceTime and Hangouts applications separately. The performance of each model is presented in terms of precision, accuracy and recall for three applications after labeling stage. We present micro-average metrics of accuracy, precision and recall, as the macro-average does not yield class importance~\cite{micro-avg}. Micro-average aggregates the contributions from all the classes to compute the average metric. We also report the mean squared error (MSE) for all models. We observe at least 88\% accuracy for all the models in all applications, except KNN regressor. This discrepancy is due to the weakness of KNN with multiple features. We observe that KNN does not generalize our dataset well and predicts most of the samples incorrectly. The mis-prediction is due the fact that each sample in higher dimension is an outlier as the distance metric (euclidean distance in our model) becomes weak with more features \cite{pestov2013k}, unless the data is well separated in all dimensions.
Among all models, ADT has consistent and better accuracy in all applications with a maximum accuracy of 92\% in Skype, 90\% in FaceTime and 93.33\% in Hangouts. 
We also use boosting with other models, but boosting did not improve the accuracy.
Therefore, as ADT performs better than other models, we use this model for all the other evaluations unless otherwise specified. The best hyper-parameters for ADT from grid search are: \texttt{n\_estimators = 10, learning\_rate = 0.1} and \texttt{linear} loss \cite{ensemblemethods2008}.

\begin{figure}[t]
      \centering
      \includegraphics[width=\linewidth]{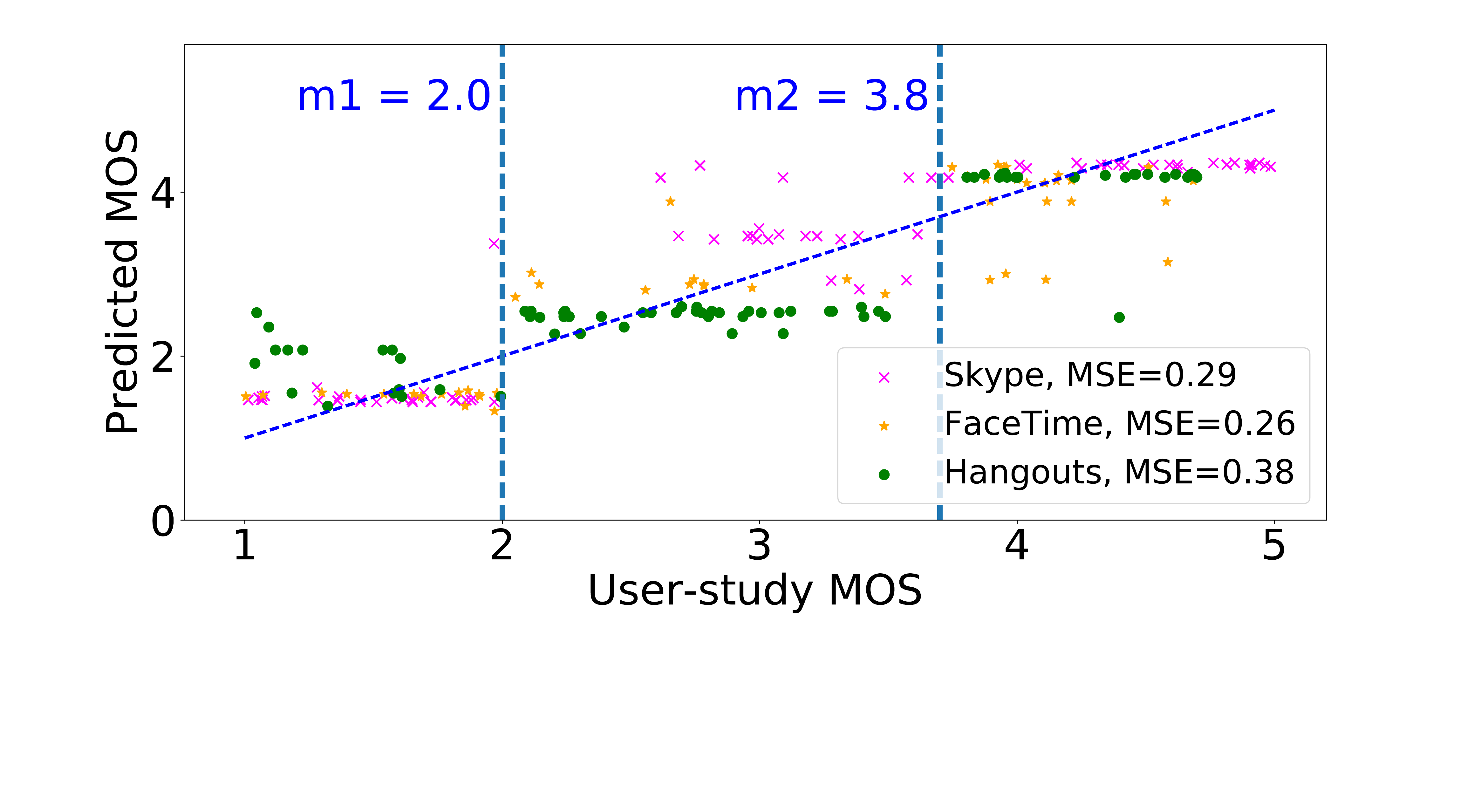}
      \vspace{-5em}
      \caption{User-study vs. predicted MOS for the three applications. We also show 3 clear QoE clusters divided by our thresholds $m_1=2$ and $m_2\approx 3.8.$}
      \vspace{-0.4cm}
      \label{fig:scores}
\end{figure}

Fig. \ref{fig:scores} shows a scatter plot of user-study MOS vs.~predicted MOS for the three applications. The MSE in MOS for the three applications is smaller than $0.4$ with ADT for all the applications. We also observe three clear clusters approximately divided by the MOS scores $m_1=2$ and $m_2=4$, that coincide with our thresholds for labeling the scores into 3 classes. This also justifies our design choice to finally employ three labels (bad, average, good) out of 5 MOS scores.

\noindent\textbf{Model performance for Skype:} 
Table \ref{class-skype} shows performance of Skype model across different regressors. 
The MSE in MOS is $0.29$ in case of ADT and it is $0.63$ with KNN regressor. 
Similarly, precision, accuracy and recall for ADT is the highest, while KNN being lowest.
ADT model gives a best threshold of $m_1=2$ and $m_2=3.8$ in separating the MOS scores into labels. While all the other models produce a threshold of $m_1=2$ and $m_2=3.4$.
Here, the best performance in ADT results from $(i)$ its low MSE and $(ii)$ the wide range for average class separation, i.e., it labels all the samples from MOS $2$ to $3.8$ as average class, while other models yield average class from MOS $2$ to $3.4$.
The performance gap is due to the distribution of our average class samples spread over wider range of bad or good labels. 
Using $m_1=2$ and $m_2=3.8$ thresholds, we get 30\%, 40\% and 30\% of our 300 samples in bad, average and good labels.

To show that our Skype model is device independent, we further evaluate our model across three devices: SG-S6 phone, SG-S2 Tab and Pixel Tab. Table \ref{label:skype-devices} shows precision, accuracy and recall for all three devices. We measure performance by training on one device, and testing on other devices. We observe that the performance is always better when trained and tested on the same device compared to training on one device and testing on other device. However, we find a difference of less than $7\%$ in accuracy when trained and tested across different devices, with an accuracy of at least 83\%. This corroborates that our model is robust across devices and it can be trained and tested without device constraints i.e., our metrics can be collected on a certain device and can be applied to any other devices.

\noindent\textbf{Model performance for FaceTime:} 
Table \ref{class-facetime} shows performance of FaceTime model across different regressors.
Similar to Skype, we observe similar performance ($>89\%$ accuracy) for all models except the KNN regressor. 
Here, although the RF regressor is performing better ($90.37\%$ precision) than ADT, the MSE in MOS is larger than ADT.
Interestingly, all models produce same thresholds of $m_1=2$ and $m_2=3.4$ in labeling the scores.
Here, the samples are distributed uniformly across three classes unlike Skype, hence all regressors are performing almost equally. However, KNN regressor still suffers in FaceTime model due to weakness with many features as explained above. 
Using these thresholds, we get 30\%, 36\% and 34\% of the 200 samples in bad, average and good labels.

To validate that our FaceTime model is device independent, we train and test across iPad and iPhone devices.
Table \ref{label:facetime-devices} shows that when training and testing on same device, we observe $92\%$ accuracy. Whereas, training  and testing across devices yields at least $88\%$ accuracy. Hence, we observe a difference of $4\%$ accuracy across device training and testing. Our FaceTime model is also device-independent. Note that, we are not comparing the performance of our model training on Android devices and testing on iOS devices and vice-versa, because the recording set-up is different these environments.

\noindent\textbf{Model performance for Hangouts:} 
Table \ref{class-hangouts} shows performance of Hangouts model across different regressors.
Similar to other applications, ADT outperforms other models with 93.33\% accuracy with an average MSE of 0.38.
Similar to FaceTime, we observe that all models produce same thresholds of $m_1=2$ and $m_2=3.5$ in labeling the scores.
Using the above thresholds, we get 32\%, 34\% and 34\% of the 300 samples in bad, average and good labels respectively.

We further evaluate the Hangouts model across three devices: SG-S6 phone, SG-S2 Tab and Pixel Tab.
Table \ref{label:hangouts-devices} shows that an accuracy of at least 86.67\% when training and testing on same device, while inter-device train and test gives an accuracy of at least 82\%.
Overall, we observe less than 8\% accuracy difference when trained and tested across different devices. This shows that the model is independent of where training and tested are conducted.

\subsection{Comparison with Baseline and Feature Importance} 
\begin{figure}[t]
      \centering
      \includegraphics[width=0.45\textwidth]{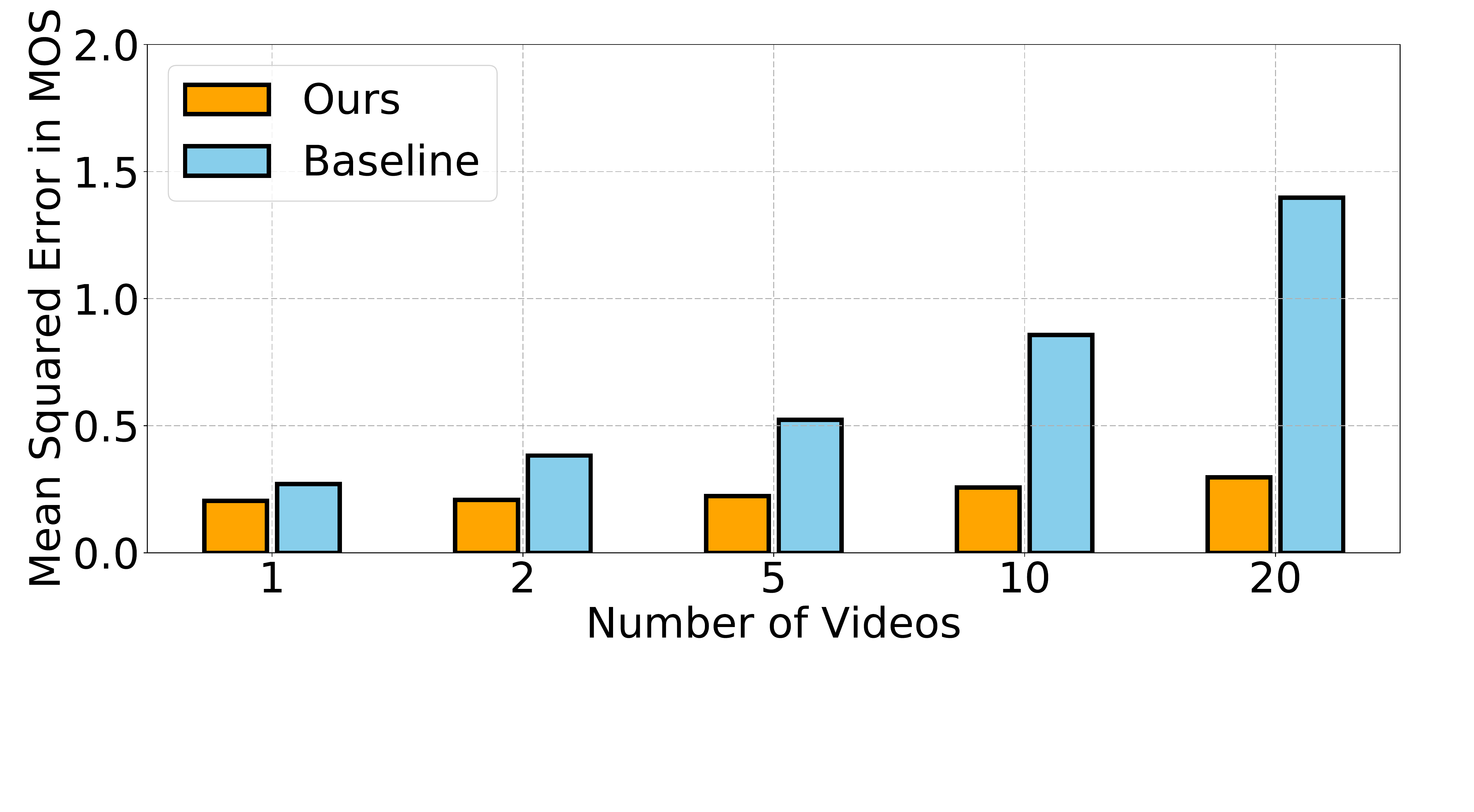}
      \vspace{-2em}
      \caption{Comparison of our metrics vs.~baseline. Baseline has an up to 1.4 MOS estimation error across 20 video contents, hence failing to distinguish good/bad from average QoE. }
      \vspace{-1.0em}
      \label{fig:baseline}
\end{figure}

We compare our model prediction error with previous work for Skype application. As we need no-reference metrics for the baseline comparison, we use the DCT blur metric used by Jana {\em et al} as spatial metric and frame-drop metric in \cite{usman2017no} as temporal metric. We fit the ADT model with these two metrics as well as with our metrics using the MOS scores from our user-study. Fig. \ref{fig:baseline} shows the performance of Skype application across the 20 videos described in Section \ref{MOTIVATION}. Clearly, for a single video, both baseline and our model yield less than 0.2 MSE in MOS whereas as the number of videos increases, the MSE grows larger than 1.3 MOS for baseline metrics. Whereas, our model has a maximum of 0.4 MSE. The baseline metric's high MOS error with large number of videos is due to DCT metric's inability to scale across diverse video content. In fact, in our experiments we measure feature importance, which is defined as the amount each feature improves performance weighted by the number of samples this feature is responsible for. We observe that feature importance for DCT is as low as $0.1$, and it is $0.9$ for frame-drop metric. Therefore, previous metrics fail to model video telephony QoE across diverse videos. We observe similar results for FaceTime and Hangouts.

\begin{figure}[t]
      \centering
      \includegraphics[width=0.45\textwidth]{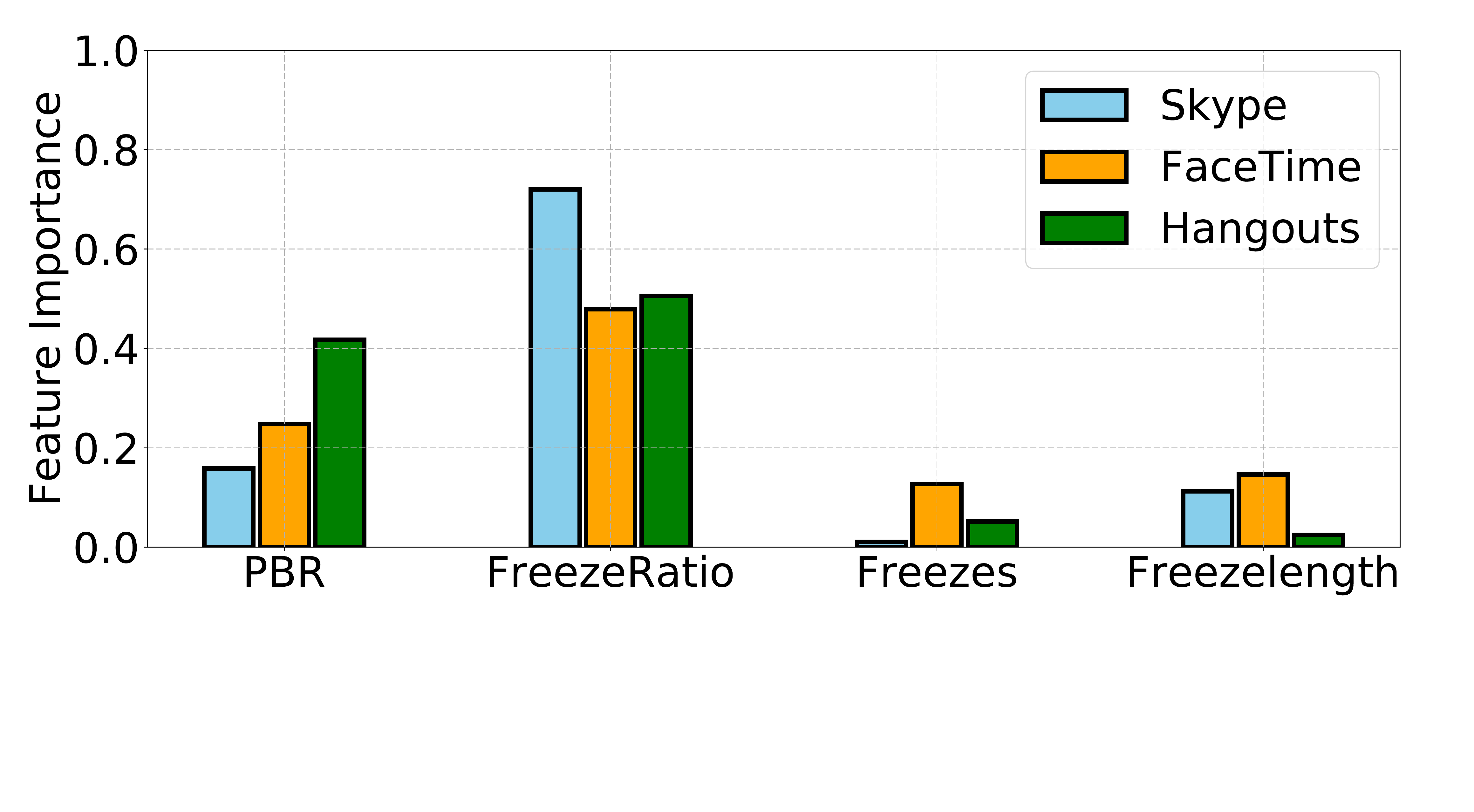}
      \vspace{-4em}
      \caption{Feature Importance for the three applications.}
      \vspace{-1.0em}
      \label{fig:ft-imp}
\end{figure}

We also evaluate the importance of each proposed metric. Fig. \ref{fig:ft-imp} shows the importance of features for ADT over all three applications. Out of all features, freeze ratio is dominating with at least $0.5$ feature importance on all applications. For Skype, we observe highest ($>0.7$) importance to freeze ratio. The rest of the metrics are almost equally important, and removal of any of these features causes an up to 0.2 accuracy degradation. Whereas, for FaceTime and Hangouts, we notice high importance for PBR ($0.28$ for FaceTime and $0.41$ for Hangouts) due to their compromise in quality over freezes.

\subsection{Summary}
The model built from handcrafted features such as PBR, freeze ratio and others has a median of 90\% accuracy in QoE labeling. However this accuracy would further reduce when applied to traditional QoS to QoE mapping model. Hence, we need to have a model that gives more accurate estimation of the labels so that the QoS to QoE mapping model will have acceptable performance. Hence, we 1) seek to explore solutions than further improve accuracy ($>$90\%), 2) use deep learning for QoE for the first time and compare with handcrafted features.
we choose recent success of deep learning solutions in accurately classifying the MOS of video samples to come up with a generic model. We explain the deep learning based video quality assessment in the following section.

\section{Deep Learning for QoE Annotation}
\begin{figure}[t]
	\centering
    \vspace{-0.4cm}
	\includegraphics[width=0.4\textwidth]{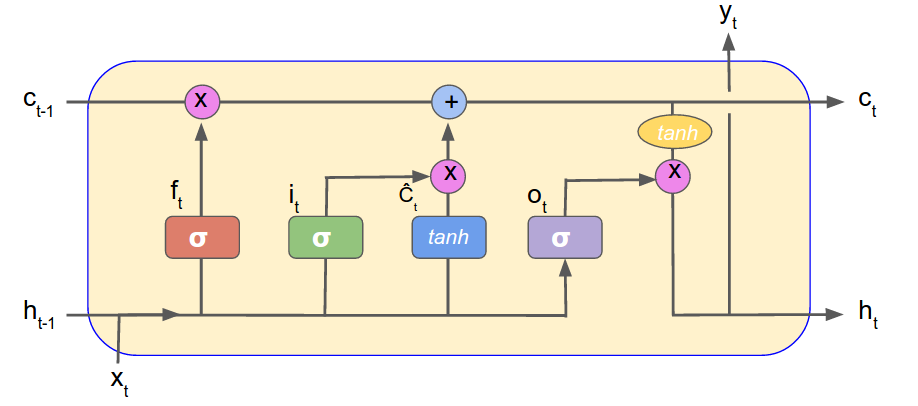}
    \caption{A Standard LSTM Architecture \cite{lstmcolah2018}.}
	\label{fig:lstmarch}
\end{figure}
\begin{figure}[t]
	\centering
	\includegraphics[width=0.4\textwidth]{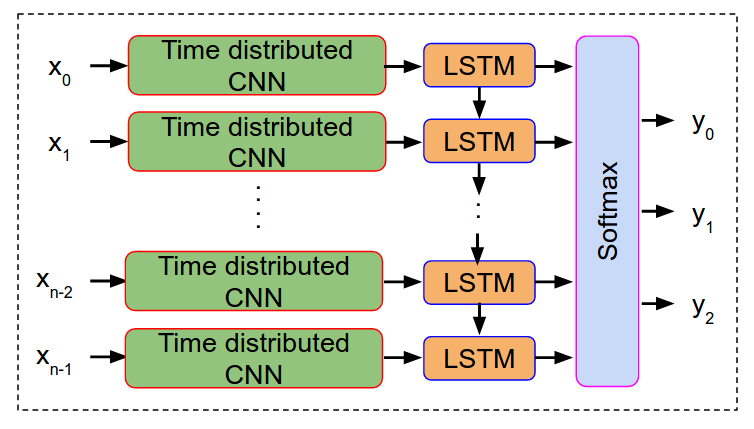}
	\caption{DCL Architecture for Video Quality Assessment.}
	\label{fig:deeparch}
\end{figure}

The evolution of parallel programming models for GPUs enables several deep-learning solutions that can be leveraged for QoE annotation \cite{wu2015modeling}\cite{ng2015beyond}\cite{liu2016spatio}\cite{tran2015learning}.
Many of these deep-learning techniques are developed for image or video classification tasks.
In this work, we leverage these techniques and propose our own model for our video QoE assessment that is agnostic to application, content and motion present in the video.
 However, the challenge here is to capture both spatial (blur) and temporal (freezes) artefacts of video while training the network. Many of the recent deep learning solutions aim for just image quality assessment which requires capturing only spatial artefacts. This is typically captured using convolutional neural networks (CNNs). To recognize and synthesize the temporal dynamics of videos, we exploit the use of long short term memory (LSTM) which keeps track of long term dependencies \cite{wu2015modeling}\cite{liu2016spatio}\cite{tran2015learning}.
A standard LSTM architecture is shown in Fig. \ref{fig:lstmarch}. 
The first step of an LSTM model is to reduce unnecessary information, which is usually called \texttt{forget layer, $f_t$} that looks for $h_{t-1}$ and $x_t$ as shown in Equation (1). Next, the model has to identify what new information to store in the cell state which is called as \texttt{input gate layer} as shown in Equation (2) and (3). The old cell state is updated into the new cell state by using Equation (4). Finally, based on the state of the cell the output is given using Equations (5) and (6).
We extract sequence of images: \{$x_0, x_1...,x_{n-1}$\}, from the video clip and feed them to a time distributed CNN followed by an LSTM to store the time-domain information. The LSTM model outputs the probability of these sequences together to determine the quality of the video clip. This can be represented by the Equation (7), where \texttt{w} represents the weights of the model after training process. A step-by-step guide and more details of LSTM algorithms is given in \cite{lstmcolah2018}.

\begin{equation}
    f_t = \sigma (W_f.[h_{t-1}, x_t]i + b_f)
\end{equation}
\begin{equation}
    i_t = \sigma (W_i.[h_{t-1}, x_t] + b_i)
\end{equation}
\begin{equation}
    \tilde{C}_t = tanh (W_C.[h_{t-1}, x_t] + b_C)
\end{equation}
\begin{equation}
    C_t = f_t*C_{t-1}+u_t*\tilde{C}_t
\end{equation}
\begin{equation}
    o_t = \sigma (W_o.[h_{t-1}, x_t] + b_o)
\end{equation}
\begin{equation}
    h_t = o_t*tanh (C_t)
\end{equation}
\begin{equation}
  videoQaulity  = \argmax_{ {0\leq i\leq 2}} P(y_i | x_0, x_1, . ., x_{n-1}, w)
\end{equation}

The proposed deep CNN and LSTM (DCL) framework is shown in Fig. \ref{fig:deeparch}. Each input image is fed to a time distributed CNN \cite{timedcnn}\cite{donahue2015long} followed by series of LSTMs. The weights of both CNN and LSTM are shared across time to scale the result across sequence of images. After the feature extraction phase from CNN and LSTM, both the spatial and temporal features are fed to fully connected network followed by a softmax classifier \cite{hinton2009replicated}. 

However, we face the challenge that any ML algorithm requires sufficient data to train the model accurately without much bias and variance. In fact, it is well known that the more data an ML model is trained, the more efficient it is \cite{perez2017effectiveness}. This is a non-trivial issue given the fact that there is no such huge dataset to evaluate video quality assessment while there are plenty of vertical video classification datasets such as YouTube 8M video dataset \cite{abu2016youtube}. This brings us to explore new techniques to enlarge our small dataset to a large-scale dataset, which is commonly referred to as \texttt{data augmentation} \cite{perez2017effectiveness}. In particular, we will train our own small net to perform a rudimentary classification. We will then proceed to use typical data augmentation techniques, and retrain our models.

\begin{table}[t]
    \centering
    \label{fig:mobilenet-arch}
    \begin{tabular}{l|l|l}
        \hline 
        \cline{1-3}
        \multicolumn{1}{l|}{\textbf{Type/Stride}} & \textbf{Filter Shape} & \textbf{Input Size} \\ \hline \cline{1-3}          
        Conv/s2 & 3 $\times$ 3 $\times$ 3 $\times$ 32 & 224 $\times$ 224 $\times$ 3 \\
        Conv dw/s1 & 3 $\times$ 3 $\times$ 32 dw & 112 $\times$ 112 $\times$ 32 \\
        Conv/s1 & 1 $\times$ 1 $\times$ 32 $\times$ 64 & 112 $\times$ 112 $\times$ 32 \\
        Conv dw/s2 & 3 $\times$ 3 $\times$ 64 dw & 112 $\times$ 112 $\times$ 64 \\
        Conv/s1 & 1 $\times$ 1 $\times$ 64 $\times$ 128 & 56 $\times$ 56 $\times$ 64 \\
        Conv dw/s1 & 3 $\times$ 3 $\times$ 128 dw & 56 $\times$ 56 $\times$ 128 \\
        Conv/s1 & 1 $\times$ 1 $\times$ 128 $\times$ 128 & 56 $\times$ 56 $\times$ 128 \\
        Conv dw/s2 & 3 $\times$ 3 $\times$ 128 dw & 56 $\times$ 56 $\times$ 128 \\
        Conv/s1 & 1 $\times$ 1 $\times$ 128 $\times$ 256 & 28 $\times$ 28 $\times$ 128 \\
        Conv dw/s1 & 3 $\times$ 3 $\times$ 256 dw & 28 $\times$ 28 $\times$ 256 \\
        Conv/s1 & 1 $\times$ 1 $\times$ 256 $\times$ 256 & 28 $\times$ 28 $\times$ 128 \\
        Conv dw/s2 & 3 $\times$ 3 $\times$ 256 dw & 28 $\times$ 28 $\times$ 256 \\
        Conv/s1 & 1 $\times$ 1 $\times$ 256 $\times$ 512 & 14 $\times$ 14 $\times$ 256 \\
        5 $\times$ Conv dw/s1 & 3 $\times$ 3 $\times$ 512 dw & 14 $\times$ 14 $\times$ 512 \\
        5 $\times$ Conv/s1 & 1 $\times$ 1 $\times$ 512 $\times$ 512 & 14 $\times$ 14 $\times$ 512 \\
        Conv dw/s2 & 3 $\times$ 3 $\times$ 512 dw & 14 $\times$ 14 $\times$ 512 \\
        Conv/s1 & 1 $\times$ 1 $\times$ 512 $\times$ 1024 & 7 $\times$ 7 $\times$ 512 \\
        Conv dw/s1 & 3 $\times$ 3 $\times$ 1024 dw & 7 $\times$ 7 $\times$ 1024 \\
        Conv/s1 & 1 $\times$ 1 $\times$ 1024 $\times$ 1024 & 7 $\times$ 7 $\times$ 1024 \\
        Avg Pool/s1 & Pool 7 $\times$ 7 & 7 $\times$ 7 $\times$ 1024 \\
        FC/s1 & 1024 $\times$ 1000 & 1 $\times$ 1 $\times$ 1024 \\
        Softmax/s1 & Classifier & 1 $\times$ 1 $\times$ 1000          \\ \hline 
    \end{tabular}
    \caption{ Mobilenet Architecture used in our DCL model. $CNN_1$ comprise layers till Avg-Pool/s1 while $CNN_2$ comprise layers from FC/s1 till Softmax. Conv-dw indicates depth-wise separable convolution.}
    \vspace{-0.4cm}
\end{table}
\subsection{Data Augmentation }
Data augmentation has shown significant benefits in training deep neural networks. There are three approaches to data augmentation based on recent advances in computer vision (CV) techniques \cite{perez2017effectiveness} \cite{tran2017bayesian}. 

\noindent \textbf{Frame Interpolation:} This method is to produce more data before the model is being trained. This is a traditional CV method where an image is applied different transformations such as rotation, translation and scale to replicate the ground-truth data. All the augmented data is fed into the model while training and only part of the original data is used for testing and validation. In this method, we can take the samples from original data and manipulate the features to generate additional data. In our dataset, we have video samples of 30 seconds each with 30 fps, which is a total of 900 frames each video. While training, we sub-sample the number of frames to minimize the storage requirements, which results in loss of information. We exploit this sub-sampling method to generate multiple samples out of a single sample. That is, while extracting frames from the video, we sub-sample the video with in multiple positions so that each sample data contain neighboring frames. For example, we can extract a frame for every 33 ms. We generate a sample by sliding the starting position of first frame with a gap of 33 ms. This way, we can generate 32 such samples for each video sample with 30 times reduction in number of frames in a sample. With this, we get a total of 25600 samples (32 times our dataset of 800 samples).

\noindent \textbf{Learning the Augmentation:} This method is to learn the data augmentation using a pre-neural network which is placed in front of actual classifying neural network. The prepended network generates new data by combining features of two samples in the original dataset. This newly generated and the original data together go into subsequent classifying layer-network. The training loss is back-propagated to classifying network as well as all back to prepended network. In our method, we can feed two video samples to the augmentation network which generates a new sample which is same size as the original samples. The augmented as well as the original samples are given as input to the classifying network. After training each sample, the classification loss and the augmentation loss is collected. The final loss is considered to be the weighted average of the two losses. 

\noindent \textbf{Generative Adversarial Networks}: This method takes an image or video as input and applies a set of styles from a subset of 6 styles--- Monet, Ukiyoe, Cezanne, Enhance, Van Gogh and Winter. A new style-transformed representation of original input is created and both original and styled objects are fed to train the network. The data augmentation using GANs and style transfer can be found in \cite{yang2017uncertainty}.

In our study, we are constrained to not alter the video sample. This makes the latter two techniques not suitable for our work in which case the generated content will have altered video qualities than the original samples. Therefore, we choose to opt frame interpolation technique and train our model using the 25600 samples as described before. 
\begin{figure}[t]
	\centering
    \vspace{-0.4cm}
	\includegraphics[width=0.8\linewidth]{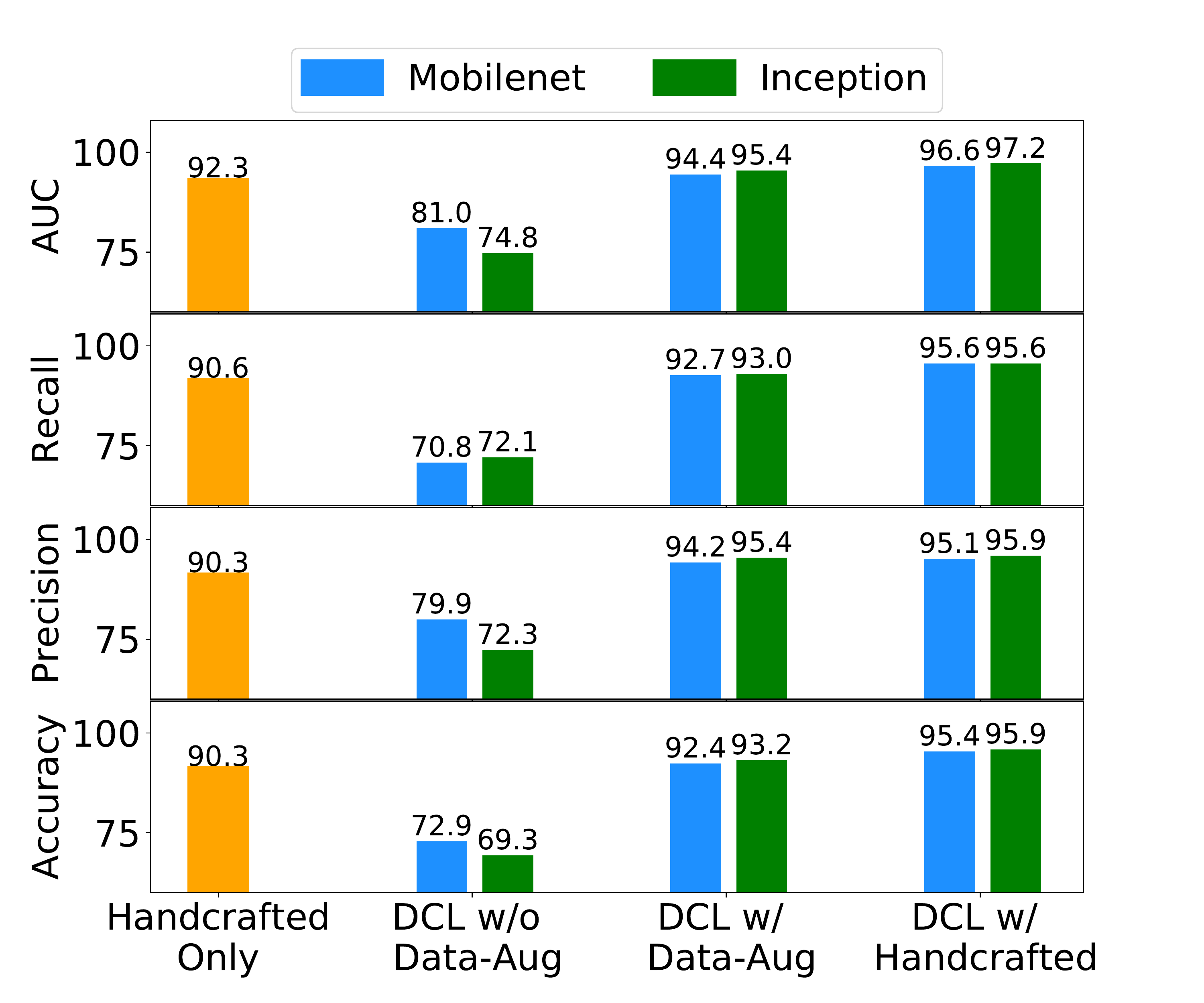}
    \vspace{-0.4cm}
	\caption{Performance of Mobilenet and Inception models. The average precision, accuracy, recall and AUC of Mobilenet on par with Inception, while the features are 10 times less than the latter.}
	\label{fig:deepperf}
\end{figure}

\subsection{Feature Augmentation }
Deep learning has proven to provide greater benefits compared to traditional ML models that work on handcrafted features in many CV applications. However, recent research in CV shows that deep neural networks learn very different latent features compared to handcrafted features \cite{guo2017locally}. This enabled many applications to combine both handcrafted features and CNNs to form a hybrid model in classification \cite{nguyen2018combining}\cite{guo2017locally}. Taking this as inspiration, we propose a hybrid DCL architecture which combines our visual features from CNN-LSTM model and the handcrafted features such as PBR and freeze ratio (presented in \S \ref{label:design}). We feed the features collected from CNN-LSTM and handcrafted features to Softmax for final classification.
\subsection{ Evaluation }
We implement our DCL architecture in keras using tensorflow libraries in Jupyter Notebook \cite{kerast}\cite{jupnt}. To extract the visual features, we exploit transfer learning techniques to inherit weights from pre-trained models. We use different pre-trained models on ImageNet \cite{deng2009imagenet} such as ResNet18, VGG16, Inception and MobileNet \cite{canziani2016analysis} to extract the spatial features. In our experiments, we observe that pre-trained and fine-tuned models are able to transfer the visual features learned from ImageNet dataset to accurately to identify quality of videos. We find similar accuracy results across all the pre-trained models and hence present the results for only Mobilenet and Inception \cite{szegedy2017inception}. We specifically focus on evaluating our model using Mobilenet pre-trained model because it has almost 10 times less features compared to other models, while providing same performance. Being lightweight, Mobilenet reduces the training significantly. Mobilenet uses depth-wise separable convolution technique \cite{chollet2016xception} and drastically reduces the dimensions. More details of Mobilenet can be found in \cite{howard2017mobilenets}.

We use the hardware configuration of Intel(R) Xeon(R) CPU E5-2603 v3 @ 1.60GHz x 2, Nvidia GeForce GTX 1070 GPU, 16 GB RAM, 500 GB Disk space and 256 GB SSD. It took 2 hours to train the proposed model and less than a second for classifying the quality of each video. This shows that even though our model is a bit complex during training phase which is usually conducted offline, the online testing is very fast and the network administrator does not have spend much time in labeling the videos.

\subsection{Results}
We evaluate our model performance in terms of accuracy, precision, recall and Area under the curve (AUC) as described in \S \ref{label:model}. We present the results by taking median for each metric on a 10-fold cross validation process.

\begin{figure}[t]
	\centering
    \vspace{-0.2cm}
	\includegraphics[width=0.8\linewidth]{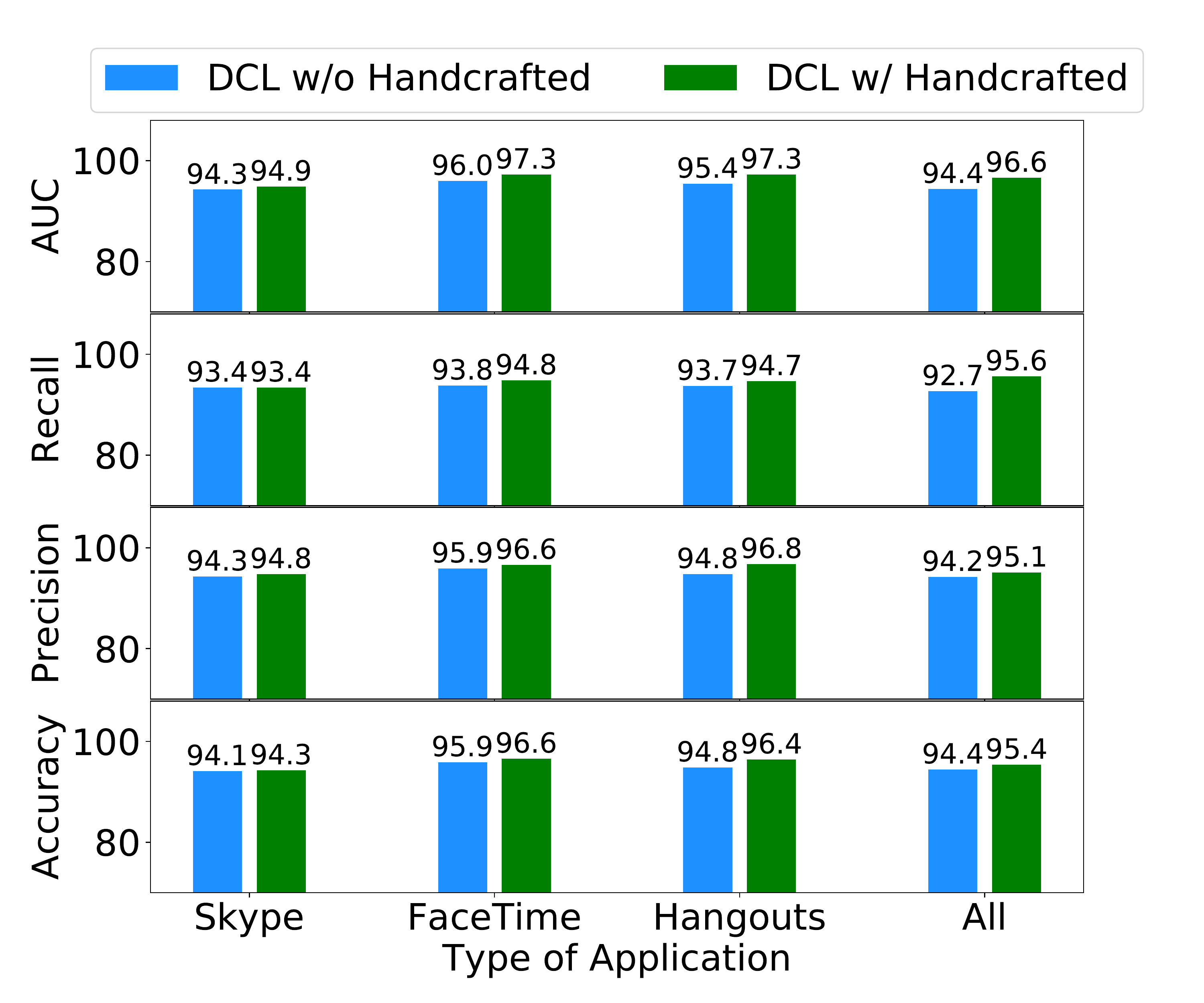}
    \vspace{-0.4cm}
	\caption{Individual model performance of applications}
    \vspace{-0.4cm}
	\label{fig:deepapps}
\end{figure}

Fig. \ref{fig:deepperf} shows the performance of our model with handcrafted features described in \S \ref{label:model} as well as deep learning classification. 
We observe a median 93\% accuracy as well precision with our DCL model compared to handcrafted model.
This is because our DCL model captures spatial and temporal features hidden from handcrafted features. As an example, we observe different types motion flows in LSTM filters that are fed as high level features to our fully connected network.
We observe similar performance with other pre-trained models such as Inception and ResNet \cite{canziani2016analysis}.
However, these models have significantly higher (about 10$\times$) parameters compared to Mobilenet as described above, henceforth we proceed with Mobilenet for the rest of our evaluations.
The accuracy goes up to almost 95\% with our hybrid approach where the model learns both handcrafted features as well as deep learning based features.

We further evaluate the performance of our model with respect to individual applications.
Fig. \ref{fig:deepapps} shows the model performance across all applications and all together.
We find similar performance about median accuracy of 93\% with our DCL model and about 94\% accuracy with hybrid model for all the applications.
Even though, individual models are performing better, the combined generic model with all the data together has very good performance with less than 1\% degradation in accuracy.
This is especially important for network administrators because there is no longer any need to identify type of application. Instead, the generic DCL model can be deployed in real-time as we reduce the training overhead.
Interestingly, we find almost no improvement in model performance for Skype by adding handcrafted features while Hangouts and FaceTime have 2 to 3\% median improvement.
This discrepancy is because Skype is very conservative in image quality and compromises temporal quality. However, motion flow is captured by our DCL model but is not able to get new spatial information from handcrafted features. This explains the reason behind no improvement of accuracy by augmenting handcrafted features to our DCL model.

\begin{figure}[t]
	\centering
    \vspace{-0.2cm}
	\includegraphics[width=0.8\linewidth]{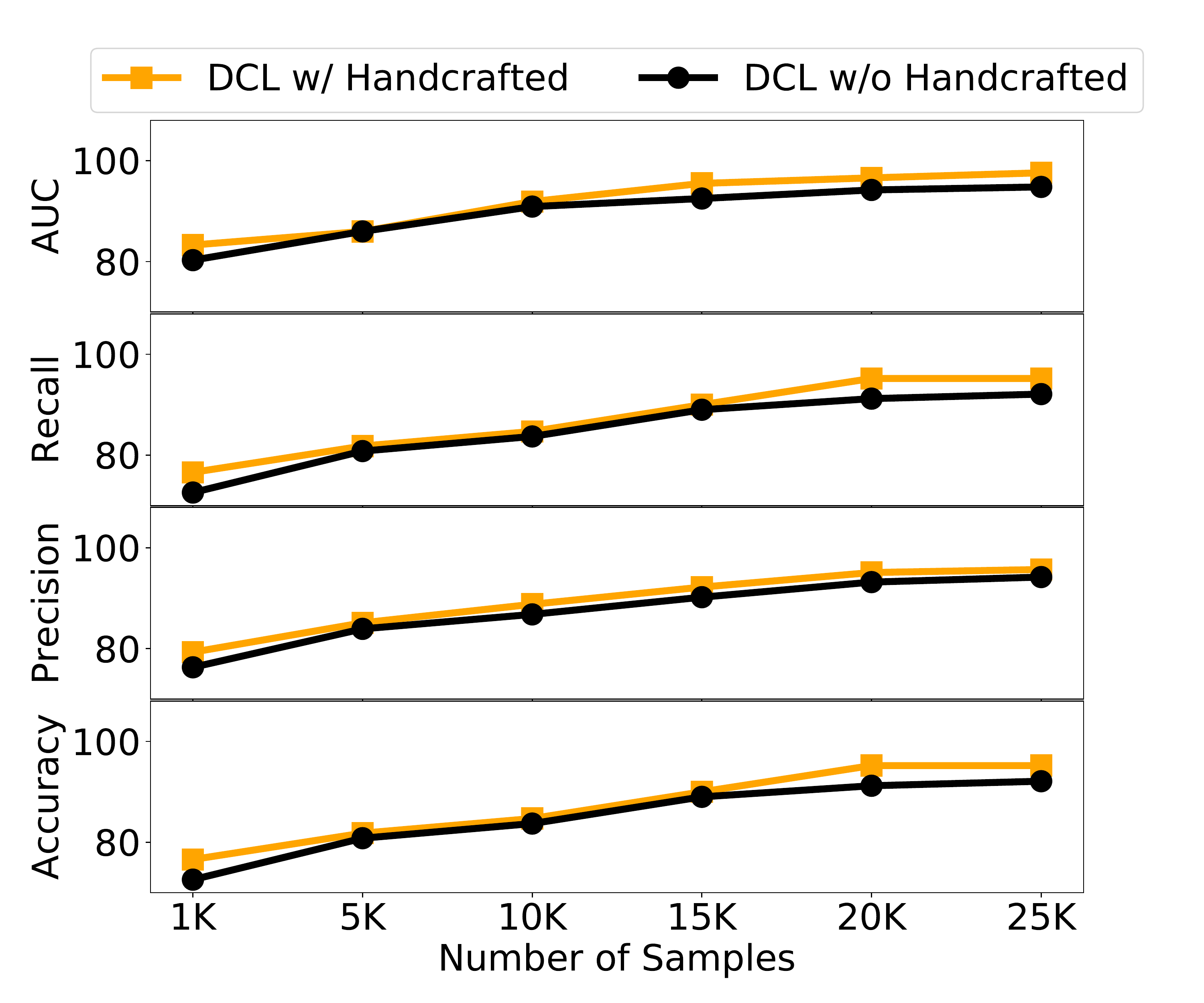}
    \vspace{-0.4cm}
	\caption{Model performance vs. number training samples.}
    \vspace{-0.4cm}
	\label{fig:deepsamples}
\end{figure}

However, the above performance is trained under rich set of data generated from our original samples. 
In practice, the model can overfit the data when trained with large amount of data.
Moreover, it is not always possible to get such large amount of data.
To this end, we investigate the performance our model with different number of samples. 
Fig. \ref{fig:deepsamples} shows the model performance with training data size increased in steps of 5000 samples.
While less than 15K samples the model performs worse than pure handcrafted based model (\S \ref{label:model}), after 20K the model does not any improvement.
This shows that our model can effectively learn the features and fit well with 20K samples and does not require more than that.
This also verifies that our model does not overfit the data, as adding more samples does not degrade accuracy.

\section{Conclusion}
Enterprise network administrators need QoE information to model QoS to QoE relationship and to efficiently provision their resources. Currently, applications do not provide QoE ground-truth. In this work, we address this problem for video telephony by introducing four scalable, no-reference QoE metrics that capture spatial and temporal artefacts of video quality. We investigate the performance of our metrics over three popular applications -- Skype, FaceTime and Hangouts -- across diverse video content and five mobile devices. Finally, we map our metrics with a large-scale MOS user-study and show a median accuracy of 90\% in annotating the QoE labels according to MOS. Our metrics outperform state-of-the-art work while capturing exact user rating. We also extended our handcrafted features with deep learning based features and achieve a median 95\% accuracy across all applications. We plan to extend this study to video streaming, such as YouTube and Netflix, and VR/AR applications.

\bibliographystyle{unsrt}
\bibliography{qos-qoe}
\end{document}